\documentclass[preprint]{aastex}

\newcommand{\etal}{{\it{et al.}}~}

\newcommand{\eg}{{\it{e.g.}}}
\def\gtaprx{\, \buildrel > \over \sim \,}

\newcommand{\chisq}{${\chi}^2$~}

\begin{document}

\title{Measuring Sizes of Marginally Resolved Young Globular Clusters with HST
\footnote{Based
on observations with the NASA/ESA {\it Hubble Space Telescope},
obtained at
the Space Telescope Science Institute, operated by AURA Inc under
contract
to NASA}}

\author{
Matthew~N.~Carlson\altaffilmark{2}
and
Jon~A.~Holtzman\altaffilmark{3}
}
\altaffiltext{2}{1571 Warburton \#5, Santa Clara, CA 95050, mcarlson@corsair.com}
\altaffiltext{3}{Department of Astronomy, New Mexico State
University, Dept 4500
Box 30001, Las Cruces, NM 88003, holtz@nmsu.edu}

\begin{abstract}

We present a method to derive sizes of marginally resolved star
clusters from HST/WFPC2 observations by fitting King models to
observations.  We describe results on both simulated images and on
observations of young compact clusters in NGC 3597 and NGC 1275.  From
the simulations, we find that we can measure King model concentrations ($c$)
to an accuracy of about a factor of two for all combinations of $c$ and
King radius ($r_0$) of interest if the data have high S/N ($\gtaprx 500$ for 
the integrated brightness).  If the concentration is accurately measured, we
can measure the King radius accurately. For lower S/N,
marginally resolved King profiles suffer from a
degeneracy; different values of the concentration give different 
$r_0$ but have comparable
reduced \chisq values. In this case, neither the core radius nor the
concentration can be constrained individually, but the half-light
radius can be recovered accurately.

In NGC 3597, we can only differentiate between concentrations for the
very brightest clusters; these suggest a concentration of $\sim$ 2.
Assuming a concentration of 2 for the rest of the objects, we find an
average King radius for the clusters in NGC 3597 of
0.7 pc, while the clusters in NGC 1275 have an
average radius of 1.1 pc. These are similar to the
average core radii for Galactic globular clusters,
0.92 pc.  We find average half-light radii of 5.4 pc
and 6.2 pc for the young clusters in NGC 3597 and NGC 1275,
respectively, while the average half-light radii of Galactic globulars
is 3.4 pc. The spread in the derived radial parameters in each cluster
system is comparable to that observed in the Galactic globular cluster
system.

\end{abstract}

\keywords{galaxies: star clusters}

\section{Introduction}

In the past several years, many observations of young, massive, compact
star clusters have been reported in a variety of different galaxies 
(see \eg, Ashman \& Zepf 1998).
In many respects, these objects are similar to globular
clusters as they would have appeared at a younger age. As such, they
may provide clues about the origin of globular cluster systems and the
reasons why such systems differ between galaxies of different
morphological types.

The range of masses of the young clusters as inferred
from colors and population synthesis models are within the  range of
masses of Galactic globular clusters, if one assumes an IMF that is
not deficient in low mass stars. The sizes of the clusters are somewhat
more difficult to constrain, given the distances of the galaxies in
which many blue cluster systems are found, and it is this problem that
we address in this paper.

Measuring structural parameters of young compact clusters is also
relevant to determine the degree to which star cluster destruction is
important in the evolution of galaxies. The luminosity functions of
young cluster systems studied in detail to date differ from those of
old globular clusters, as they have an exponentially increasing
distribution rather than the log-normal distribution of old globular
clusters (\eg, Carlson \etal 1998, 1999, Whitmore \etal 1999, Zepf \etal
1999 ).  As a result, cluster destruction must be very important, and
dependent on mass, if the observed young cluster systems are to evolve
to look like older systems. The alternative is that the initial mass
spectrum of clusters is different in locations today where we are
seeing them than it was in the locations where globular clusters
formed. Understanding the importance of cluster destruction is also critical
to understanding the relative number of cluster vs. field stars formed
in a merger, which leads to an understanding of whether the specific
frequency of a cluster population can change during an event in which
massive star clusters form (\eg, Carlson \etal 1999).

To date, observational attempts to constrain the sizes of young
globular clusters have been largely made by comparing cluster aperture
magnitude differences to stellar and/or model cluster aperture
magnitude differences.  Holtzman \etal (1992) used WFPC1 data to
measure the difference between concentric 2 and 4 pixel aperture
magnitudes, $m_2-m_4$, for the young clusters in NGC 1275.
From these measurements they concluded that most of the sources were
unresolved. Whitmore \etal (1993) attempted to determine the half-light
radius, $r_{h}$, for the clusters in NGC 7252 by comparing the
$m_{0.5}-m_3$ to that same difference for model clusters.  The models
consisted of stellar images broadened by Gaussians. Later work by Holtzman
\etal (1996) with WFPC2 data used a differential $m_1-m_2$ value
computed by taking the $m_1-m_2$ for the observed clusters and
subtracting the $m_1-m_2$ for an identically centered model point
spread function (PSF).  These were compared to the same differential
$m_1-m_2$ calculated for a model PSF broadened by a set of modified
Hubble profiles of varying core radii.

Both Gaussians and modified Hubble profiles are rather poor
approximations to the surface brightness profiles of globular
clusters.  In particular, the surface brightness of a Gaussian falls
off too rapidly with radius to adequately represent globulars. Galactic
globulars are better fit by King models which are based on the 
model of a cluster of equal mass stars with
a truncated Maxwellian distribution of velocities.  
In a  King model (King 1962),
the structural properties of globular clusters are represented by
three parameters - an amplitude, from the number of stars, a core
radius $r_0$, based on the internal dynamics of the system, and a limiting
(tidal) radius $r_t$ imposed by the Galactic potential well; often, 
King models are parameterized by the King radius and the concentration, 
$c\equiv \log {r_t\over r_0}$.

King models are generally good fits to the density distribution of
Galactic globular clusters (e.g., King \etal 1968).  Later studies
extended our knowledge to the structural parameters of a handful of
globular clusters in several Local Group galaxies, including the
Magellanic Clouds (Chun 1978, Elson \& Freeman 1985, Kontizas 1984), M
31 (Crampton \etal 1985), and the Fornax dwarf spheroidal (Demers,
Kunkel, \& Grondin 1990).

With the advent of the Hubble Space Telescope, measurements of globular
cluster structural parameters were pushed to NGC 1399 in the Fornax
Cluster at 16.9 Mpc (Elson \& Schade 1994). They fit King models to 24
old globulars using the Levenberg-Marquardt method and determined $r_0$
and $c$ for each.  The improved resolution of the WFPC2 has allowed the
structural parameters of even more distant globular clusters to be
studied. Kundu \& Whitmore (1998) and Kundu \etal (1999) have modeled
the light distributions of old globular clusters in the S0 galaxy NGC
3115 and M87 using King models. Their technique uses the cumulative
light profile of clusters in several concentric annuli as a way to
measure the parameters of the underlying surface brightness profile.
For comparison, they constructed numerous template cumulative
light profiles for various combinations of centering, position in the
field, King radius, and concentration. They determined the King radius
and concentration for a given cluster based on the least squares best
fit to the cluster from within their grid of template profiles.  They
noted that they were not able to uniquely fit both the King radius and
the concentration simultaneously.

What can we learn from fitting King models to young blue clusters?
Perhaps the most important reason to fit King models is to be able to
confirm the globular nature of these clusters by comparing their radial
parameters to those of known globular clusters, such as those in the
Galaxy. Another valuable piece of information comes from the link
between cluster concentrations and cluster disruption through mass loss
(Weinberg 1993). Clusters with the highest concentration, thus the
largest ratios of tidal to King radii, are most prone to disruption by
this mechanism.  Are the young globulars found in so many galaxies
truly stable, or are they going to dissolve in the next 10 Gyr? The
concentration of globulars can also give us information about tidal
shocking (Surdin 1993). Clusters with low concentrations are subject to
disruption by tidal shocks, particularly in the inner parts of
galaxies. The tidal radii of globulars are also useful as they can help
to place limits on the mass of the host galaxy (Elson \& Schade 1994);
since the tidal radii of clusters are limited by the gravitational
field of the galaxy, they act as direct probes of the galactic mass.

The current paper investigates the measurement of cluster radial
parameters by fitting the observed two-dimensional cluster light
distributions.  The inclusion of data beyond a two or three pixel
radius can help differentiate between various models and can better
constrain the radial parameters. 
Section 2 discusses the
construction of our models of observed cluster light distributions with the
HST/WFPC2, and Section 3 discusses model results on the ability of the
fitting procedure to measure King radii and concentrations as a
function of signal-to-noise and angular size. 

In Section 4, we apply our method for measuring the radial parameters of young
clusters to the cluster systems in NGC 1275 and NGC 3597. The cluster
systems in these galaxies were chosen because each has a substantial
number of bright clusters for which HST images are available.
Each also has a number of bright clusters well away from the central
portions of these galaxies where the bright galaxy background makes
accurate fitting difficult.  The clusters in these galaxies are only
marginally resolved.  Adopting $H_0=75\ km/s/Mpc$, one pixel of the HST 
PC frame has a
size of $\sim$15 parsecs for NGC 1275 ($v=5264\ km/s$), while for NGC 3597 
($v=3485\ km/s$), one pixel is equivalent to $\sim$10
parsecs. So the average core radius of a Galactic globular cluster,
$\sim$ 1 pc, corresponds to a small fraction of a pixel.
Section 5 discusses the results.

\section{Models}

To model the observed cluster images, we convolve a WFPC2 Planetary
Camera PSF with different functions including a modified Hubble
profile, a Gaussian, and a family of King models. These functions
provide different models of the underlying stellar surface density
distribution.

\subsection{PSF}

The PSFs used to model the clusters were constructed by optical
modeling of the HST mirrors and the WFPC2 camera.  The resultant WFPC2
Planetary Camera PSF differs from an ideal diffraction limited image
because of the structure of the telescope itself, which leads to 
diffraction features in the images.  Jitter from small changes in the
pointing of the telescope broadens the PSF.  Slight variations in
focus, occurring even within a single exposure, also acts to blur the
PSF. The PSF is position-dependent, due to variable aberrations
and variation of the pupil function across the PC field of view. These
introduce a radial distortion of the PSF causing it to become more
elliptical towards the edges of the field.  All model PSFs were
constructed by a diffraction calculation which modeled these effects;
models were computed at 1/4 PC pixel resolution, which corresponds to
0.011 arcsec. 
Our models also include pixel smearing within the CCD (Burrows \etal
1995).  This pixel smearing comes from the diffusion of charge in the
CCD and is modeled by spreading the counts in a given pixel into
neighboring pixels with the following distribution,

\[
\begin{array}{rrr}
1\% & 4\% & 1\% \\ 4\% & 80\% & 4\% \\ 1\% & 4\% & 1\%
\end{array}
\]

\noindent with the central pixel retaining 80\% of the original counts;
neglecting this effect can significantly affect deduced radial parameters.
These calculations are very similar to those performed
by the standard TinyTim PSF generator distributed by the Space Telescope
Science Institute and the results are nearly identical.

To estimate the effect of PSF variations and uncertainties, we experimented
with three different PSF models.  The first was our nominal PSF with a
typical focus and for an object in the center of the field.  The second
model was created with a PSF with focus representing the maximum
deviation seen in WFPC2 images. The third set of models was appropriate
for a PSF at the edge of the field.  As discussed below, the effect of
using the different PSFs on derived cluster radial parameters was
relatively small.  

One other possible source of effective degradation of the PSF comes
from summing multiple exposures through each filter since frame to frame
variations in the pointing can result in a broadened PSF. Attempts
to measure this effect in our data sets by correlating shifts in the
centroids of individual objects in each of the exposures suggested 
that the effect on the PSF is very minor for our data sets, so it was
not modelled here.

Jitter can also broaden the image, but the effect is usually smaller
than that of the poor focus we have assumed in our tests, so we did not
include it in our models.

\subsection{Functions}

\begin{figure}
\figurenum{1}
\caption{Radial plots of six King models (solid lines,
King radii of 0.01 and concentrations of 1, 1.3, 1.7, 2, 2.3, and 2.95), a
modified Hubble profile (dashed line, core radius of 0.01), and a Gaussian
(dotted line, FWHM of 0.01).}
\plotone{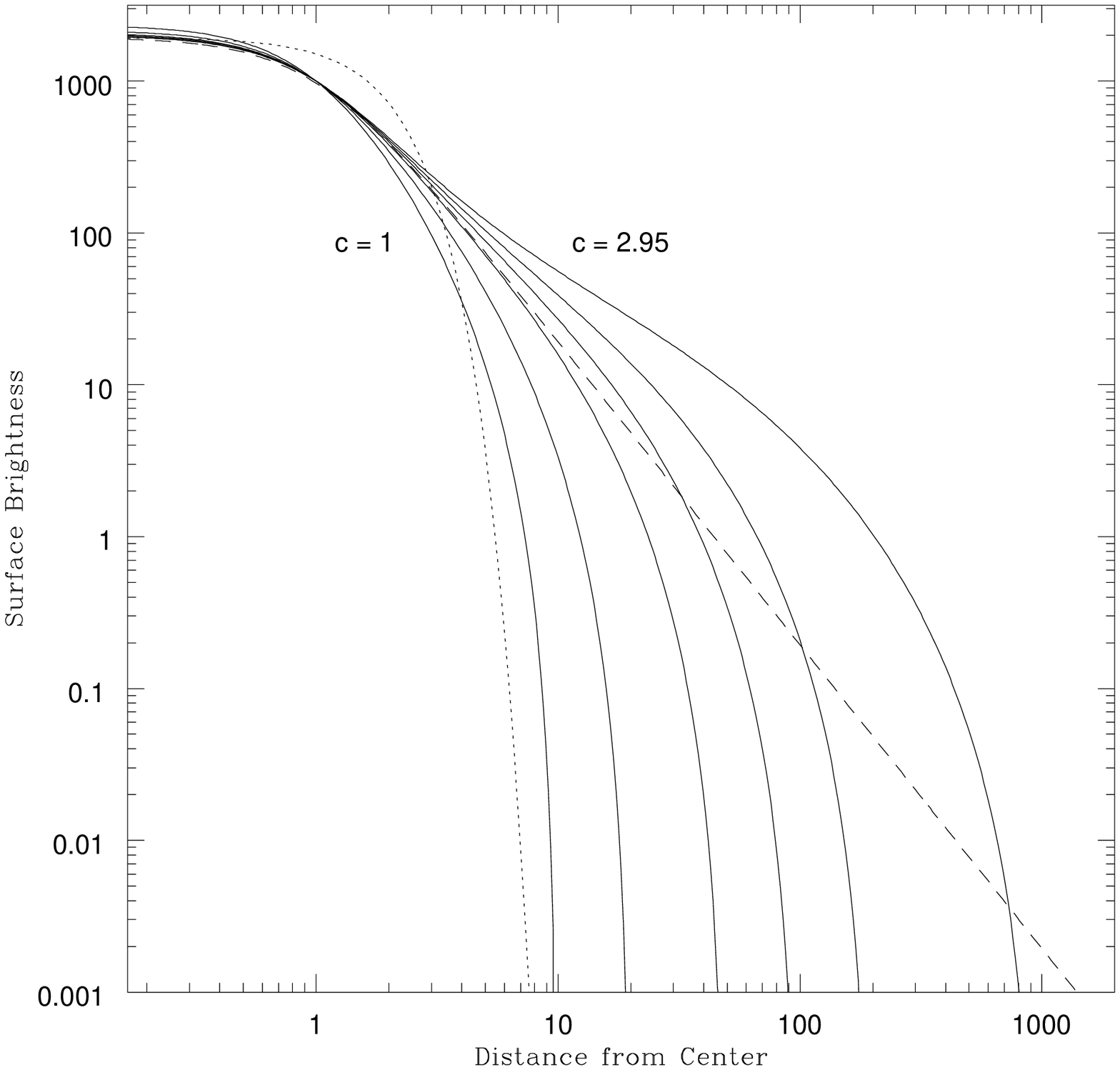}
\label{fig:functions}
\end{figure}

To simulate the surface brightness distribution of the star clusters,
we used three different functional forms: a
Gaussian (Eq. 1), a modified Hubble profile (Eq. 2), and a family of
King models. We graphically demonstrate the radial surface brightness
distribution of these functions in Figure \ref{fig:functions}.

$$I=I_0 exp(r^2/2\sigma ^2) \eqno (1)$$

$$I=I_0 \biggl(\frac{1}{1+(r/r_c)^2} \biggr) \eqno (2)$$

\noindent The King model surface brightness distributions were 
provided by C. Grillmair (private communication) and consist of a series
of King models with various concentrations. All models are normalized to
have a King radius of 1. The King radius, the tidal radius, and the
concentration are related by

$$c \equiv log(r_t/r_0) \eqno (3)$$

\noindent Note that the King radius, $r_0$, is defined by

$$r_0 \equiv \sqrt{ {9 \sigma^2} \over {4 \pi G \rho_0}} \eqno (4)$$

\noindent where $\sigma$ is the projected velocity dispersion and $\rho_0$
is the central density. The King radius and core radius (where the surface
brightness drops to one half of the central value) of King models are
very similar, and the King radius is often simply referred
to as the core radius of the King model in the literature.

As can be seen in Figure \ref{fig:functions} the Gaussian model is quite different from
the modified Hubble profile and King models. The Gaussian falls off
much more rapidly than the other models and is a rather poor fit to the
extended light profile of a globular cluster; we consider it, however, because
it has
been used by several authors to model the light distribution of young
globulars (Whitmore \etal 1993, Whitmore \& Schweizer 1995, Schweizer
\etal 1996). The modified Hubble profile provides a somewhat better fit
to the outer regions of a globular cluster surface brightness
distribution, and is perhaps the best one parameter fit.  Such a model
has been used to describe the light distribution of young globulars
(Holtzman \etal 1992, Holtzman \etal 1996, Carlson \etal 1998, Carlson
\etal 1999). While the core radius of the modified Hubble profile is
similar to the King radius, it is strictly an empirical fit to the data
with no physical significance. As discussed above, however, King models
provide the best fits to Galactic globulars and are physically
motivated.

Operationally, each of the different functions is constructed at 1/44th
of a pixel resolution and normalized to a total flux of 1 count. We
chose to construct the functions at this resolution as a compromise
between the speed of our fitting procedure and loss of accuracy.  Once
a functional form for a fit is chosen and constructed, the function is
rebinned to quarter pixel resolution to match the resolution of the
PSF.  The function and the PSF are convolved using a fast Fourier
transform, then the output model is centered and rebinned to 1 pixel
resolution.

\subsection{Fitting Algorithm}

Our fits allow for four free parameters - the $x$ and $y$ position of the
object, an amplitude, and a radial parameter.  For the Gaussian this
radial parameter is the full width at half maximum (FWHM), for the
modified Hubble profile, it is the core radius, and for the King
models, it is the King radius. King models with different concentrations
were considered independently and were compared by comparing the quality
of the fits, rather than by allowing the concentration to be a free
parameter; this avoided the need to be constantly recomputing King
models.

The fitting procedure uses the Levenberg-Marquardt method (Marquardt
1963), which is a mixture of the inverse-Hessian method and the method
of steepest descents for finding the \chisq minimum for a model
approximation of observational data.  
After each iteration, the program determines if an acceptable fit has
been reached. The fit is accepted if the change in the parameters is
less than 0.1\% of the previous values for the radial parameter and the
amplitude and less than 0.001 pixel shift in the $x$ and $y$ position of
the object and the value of $\chi^2$ has decreased in the last
iteration.  If the fitting routine doesn't converge in a few dozen 
iterations, each of the parameters is given a small displacement and
the fitting is resumed.  This aids in eliminating poor fits.  The fit
is flagged as suspect if a maximum number of iterations has been
reached. If multiple iterations occur without a decrease in $\chi^2$,
the fit is flagged and rejected.

\section{Tests of the Models}

\subsection{Effects of Signal-to-Noise Ratio}

To test the effectiveness of the fitting procedure as a function of
signal-to-noise, we constructed sets of 100 artificial clusters which
included Poisson noise in the signal plus readout noise. Each of these
artificial clusters was created using our nominal PSF and was randomly
centered within a pixel. We included in these tests three models
for the Gaussian function (FWHM $=$ 0.8, 1.4, 2.0 pixels), the modified
Hubble profile with three different core radii ($r_c =$ 0.005, 0.01,
0.03 pixels), and six King models. For the King models we chose
concentrations of 1.0, 1.3, 1.7, 2, 2.3, and 2.95, corresponding to
ratios of tidal radii to King radii of 10, 20, 50, 100, 200, and 900,
respectively.  King model clusters with $c = 2$, which provide the best
fit for the brightest of the observed clusters in NGC 3597 (discussed
below), were constructed with a set of three different radii, ($r_0 =$
0.04, 0.07, 0.15 pixels).  For the other five King models, an average
King radius was assumed for each ($r_{0,10} =0.4$, $r_{0,20} = 0.25$,
$r_{0,50} = 0.1$, $r_{0,200} = 0.03$, and $r_{0,900} = 0.005$), also
based on fits to the brightest observed clusters. We have also included
a cluster model that mimics the typical Galactic globular cluster,
with $c = 1.3$ and $r_0 = 1$ pc, equal to 0.1 pixel at the distance of
NGC 3597. 

To simplify later comparison to the observational data, we have created
the artificial clusters for each function with signal-to-noise values
which correspond to a range of magnitudes from R=15 to R=25 in steps of
1 magnitude in our data set of NGC 3597 (total exposure time of $\sim$
5000s in each filter); these correspond to S/N ratios of $\sim$ 5750, 3600,
2300, 1450, 900, 600, 400, 230, 150, 90, and 55 in the F702W filter.  
We have assumed a color that is the mean of the
blue cluster colors from our observations.  Since we have based our
simulations on the sizes of the brightest clusters in the closer
galaxy, NGC 3597, any output radial parameter corresponds to an actual
physical size which is $\sim$ 50\% larger in the more distant galaxy,
NGC 1275.  Since the background is relatively faint for many of the
observed clusters, none of these simulations include noise due to the galaxy
background, which varies depending on position; for fainter clusters
the measured spread should be taken as a lower limit.

\begin{figure}
\figurenum{2}
\plotone{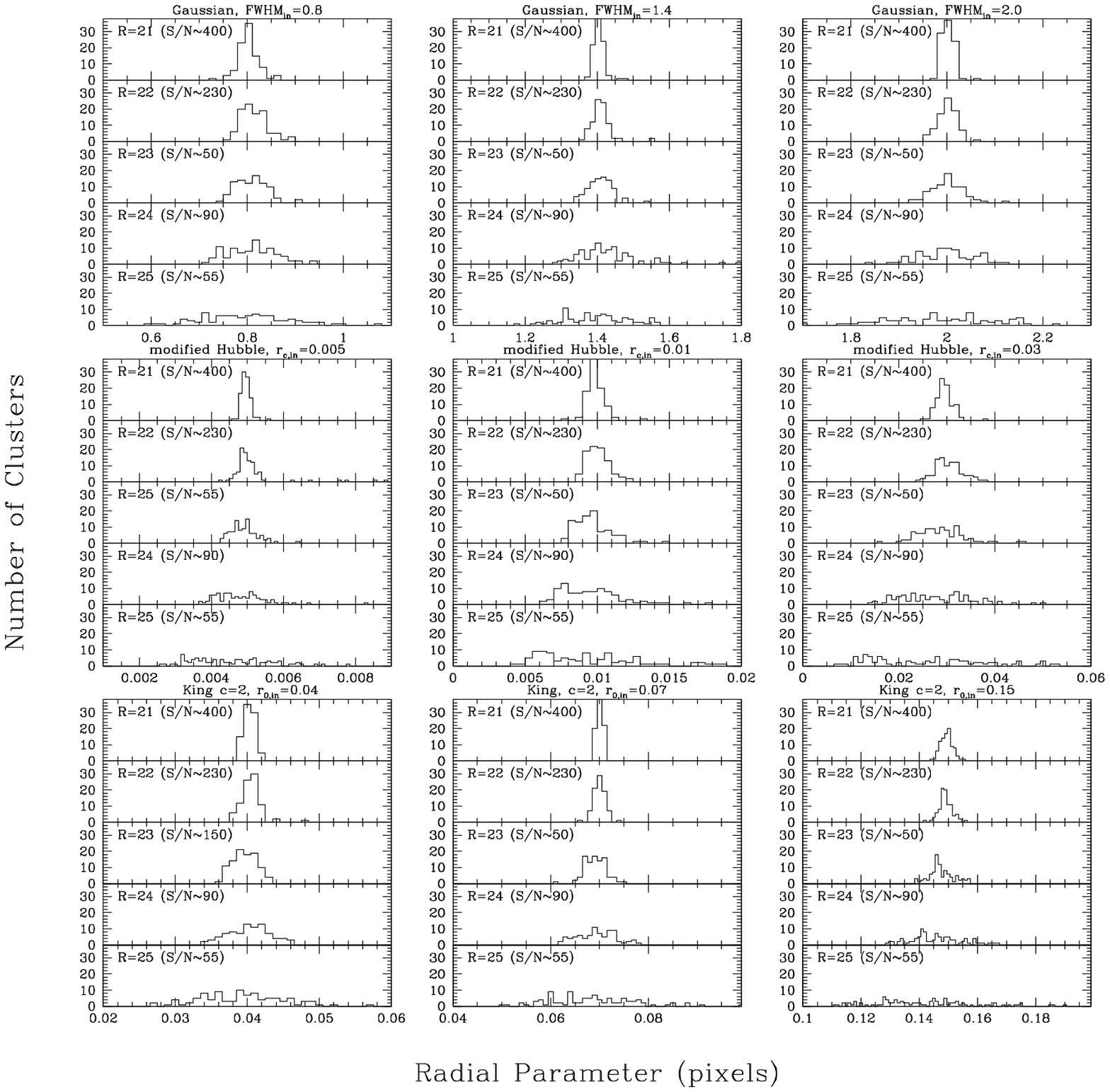}
\label{fig:radialerra}
\end{figure}

Initially, each artificial cluster was fit using the same PSF and model
function with which it was constructed, solving for the best fitting
radial parameter.  
Figure
\ref{fig:radialerra} shows histograms of the spread in the output
radial parameters for the different functions, radial parameters, and
magnitudes.  From these data we can see that if the cluster light is
well modeled by any of the functions we have chosen, the fit given by
that function will provide an estimate of the radial parameter of the
cluster which is accurate to better than $\sim$ 10\% for
signal-to-noise ratios greater than $\sim$ 100. There is possibly a
small systematic error introduced for the faintest clusters with some
of the King models, but it is not particularly large; as we will see
later, other uncertainties probably lead to larger systematic errors.

\begin{figure}
\figurenum{3}
\caption{Histograms of best fit concentration as a function of
magnitude. Input artificial cluster was given $c= 2$ and $r_0 =$ 0.07 pix,
typical of blue clusters in NGC 3597.
}
\plotone{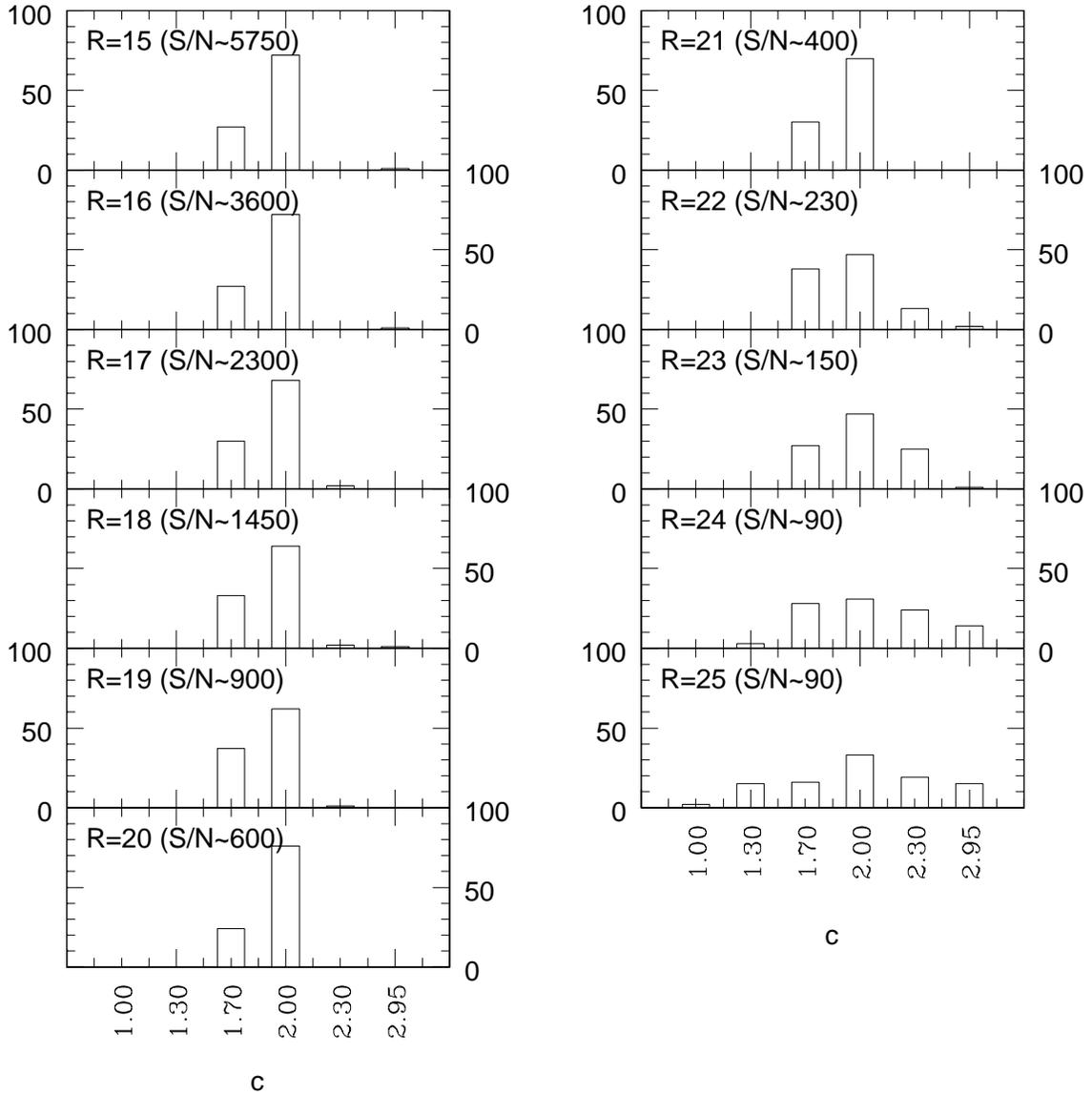}
\label{fig:concerr2}
\end{figure}

To observe the extent to which it is possible to differentiate between
different concentrations we have constructed 11 sets of 100 synthetic
clusters with signal-to-noise ratios in the range of 60 - 6000 for the
F702W PSF. For each, the input
model consists of a King model with $c = 2$ and $r_0 = 0.07$ pixels,
as motivated by our fits to the brighter observed
clusters in our NGC 3597 data, discussed below.  
We fit these using King models with different concentrations, and define
the output concentration to be the one that gives the best fit;
Figure \ref{fig:concerr2} shows histograms of the recovered
concentration.  While even for the brightest artificial
clusters there is difficulty distinguishing between concentrations of
1.7 and 2, only for signal-to-noise ratios below about 150 ($R = 23$)
do we find the correct concentration ($c=2$) in much less than about 50\%
of the artificial clusters. 

\begin{figure}
\figurenum{4}
\caption{Histograms of best fit concentration as a function of
magnitude. Input artificial cluster was given $c= 1.3$ and $r_0 =$ 0.1 pix,
similar to Galactic globular clusters if observed at the distance of NGC
3597.
}
\plotone{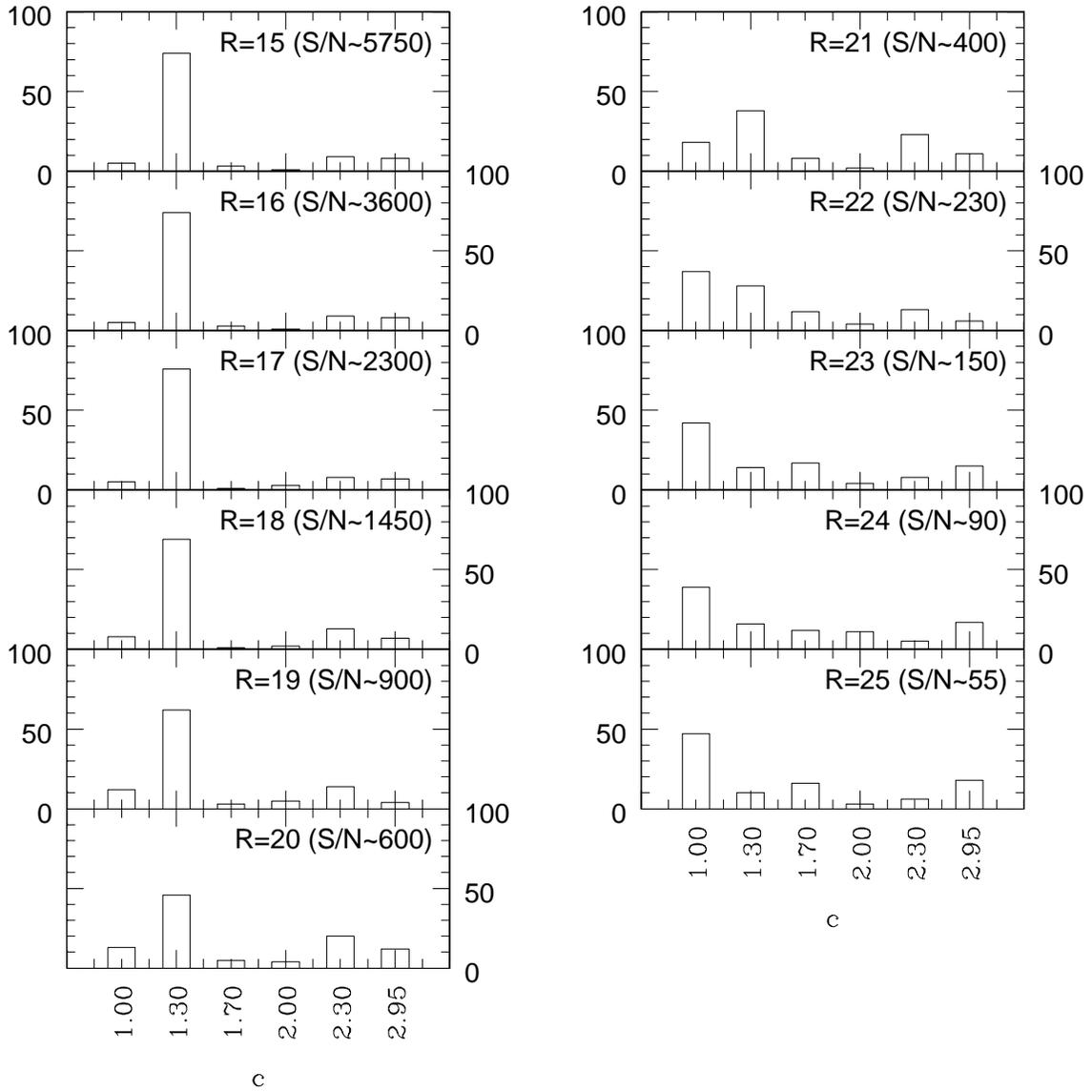}
\label{fig:concerr3}
\end{figure}

Another set of models was constructed with properties based on
Galactic globulars as they would be 
seen at the distance of NGC 3597, to see whether
we can derive concentrations for more compact
clusters.   Eleven sets of artificial clusters were created with the
same signal-to-noise range as the previous tests but with different
King model parameters ($c = 1.3$, $r_0 =$ 0.1 pix, F702W PSF).
Histograms of the best fit output concentration are shown in Figure
\ref{fig:concerr3}.  The input concentration is found to be the best
fit concentration more than 50\% of the time only above a signal-to-noise
of $\sim$ 600, reflecting the fact that constraining the concentration
is more difficult for more compact clusters. For the clusters that
mimic Galactic globular cluster structural parameters, the tidal radius
extends only two pixels from the core for this model, compared to
$\sim$ 5 pixels for the artificial clusters based on the parameters of
the brightest NGC 3597 clusters.

\begin{figure}
\figurenum{5}
\caption{Histograms of best fit concentration as a function of magnitude.
Input artificial cluster was given $c= 1.3$ and $r_0 =$ 0.28 pix, parameters
which would be similar to those expected for a Galactic globular cluster
at the distance of the Virgo cluster.
}
\plotone{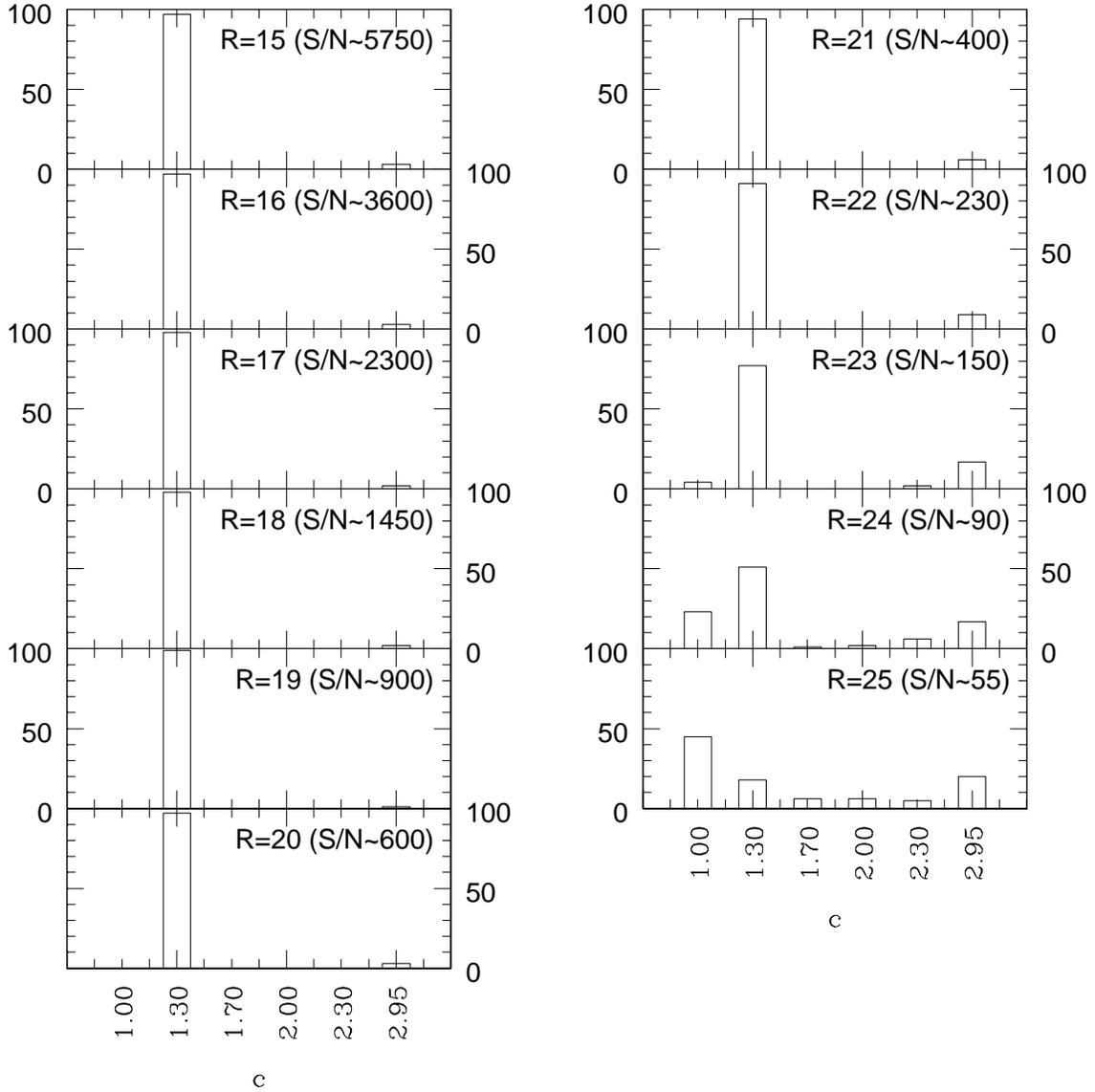}
\label{fig:concerr4}
\end{figure}

To determine how much the ability to measure radial paramters is
improved if one looks at less distant objects,
we also constructed a set of clusters which would represent the
expectation for Galactic globulars at the distance of Virgo ($c = 1.3$,
$r_0 = 0.28$ pix, F702W PSF). For these input parameters we recover the
proper concentration $\sim$ 50\% of the time down to a signal-to-noise
of $\sim$ 90. We show a histogram of the best fit concentrations as a function
of signal-to-noise in Figure \ref{fig:concerr4}.  It is clear that it
is far easier to distinguish between different concentrations in
instances where the angular size is larger.  

We note that in Figures \ref{fig:concerr3} and \ref{fig:concerr4} even
at high signal-to-noise some small number of clusters are apparently
best fit by $c = 2.3$ or 2.95.  This presumably arises because the
algorithm occasionally convererges to an incorrect local $\chi^2$
minimum. It only occurs when fitting clusters with small input concentrations.
Because it happens only rarely, we have not pursued modifications to
the algorithm to avoid the problem.

When we fit lower signal-to-noise artificial King model clusters with
King models of different concentrations, we find that the resultant
fits are, in a \chisq sense, equally valid. At lower signal-to-noise
there is a degeneracy among the family of King models which makes
uniquely determining a concentration for a given cluster impossible.
King models with larger concentrations have smaller output radial
parameters. This problem is more notable for more compact clusters.
This systematic trend can lead to errors in the output King radii that are
correlated with the chosen concentration. To avoid these errors as much
as possible, we have followed the example of Kundu \& Whitmore (1998),
who noted exactly the same effect, and chosen to use the half-light
radius, $r_h$, for cluster-to-cluster comparisons for fainter objects. This
quantity is relatively independent of the chosen combination of
concentration and King radius.

\subsection{Effect of Errors in the PSF}

All results above assumed that we have a perfect understanding of the
HST/WFPC2 point spread function.
To attempt to put some limits on the potential errors arising from an
inaccurate PSF model, we have constructed three sets of synthetic King models
using three different PSFs:
 
\begin{enumerate}
\item our nominal PSF model (PSF1)
\item a defocussed PSF (PSF2)
\item a PSF from the edge of the PC image (PSF3)
\end{enumerate}

The model clusters for this experiment had $r_0 = 0.07$, $c = 2$, and 
signal-to-noise ratios ranging from
$\sim$ 600 to 55.

We fit each set of artificial clusters using PSF1.
The artificial clusters created with PSF2 are
reasonably well fit with PSF1 models, with reduced \chisq values
differing significantly from 1 only for S/N $\gtaprx$ 400.  As might be
expected, however, the use of an incorrect PSF can lead to systematic errors.
The average output King
radii from these fits was 0.082, 17\% larger than the input
value of 0.07. Fits to clusters made with PSF3 are less 
accurate, with reduced \chisq values significantly different from 1
for S/N $\gtaprx$ 150. Still, the output best fit radii average is 0.074,
only 6\% larger than the input King radii. We conclude that while
focus and position dependent distortions may contribute to errors in
the correspondence between the theoretical profile and the
observations, they change the derived radial parameters only slightly.

\begin{figure}
\figurenum{6}
\caption{Histograms of best fit concentration as a function of magnitude.
Input artificial cluster was given $c= 2$ and $r_0 =$ 0.07 pix as in
Figure 4, but were created using PSF3 and fit using PSF1.
}
\plotone{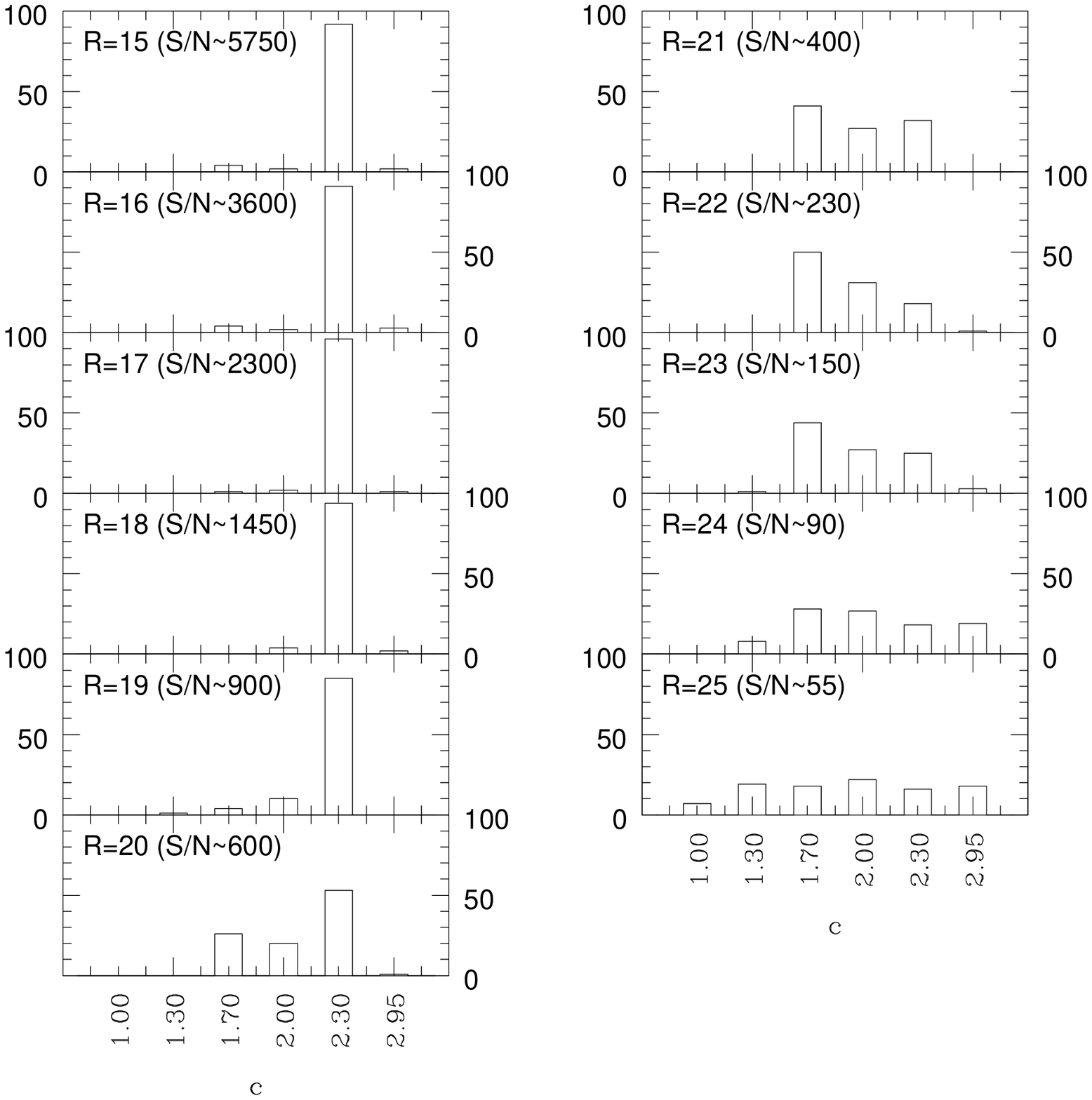}
\label{fig:psferr1}
\end{figure}

\begin{figure}
\figurenum{7}
\caption{Histograms of best fit concentration as a function of magnitude.
Input artificial cluster was given $c=1.3$ and $r_0 =$ 0.28 pix as in Figure
6, but were created using PSF3 and fit using PSF1.
}
\plotone{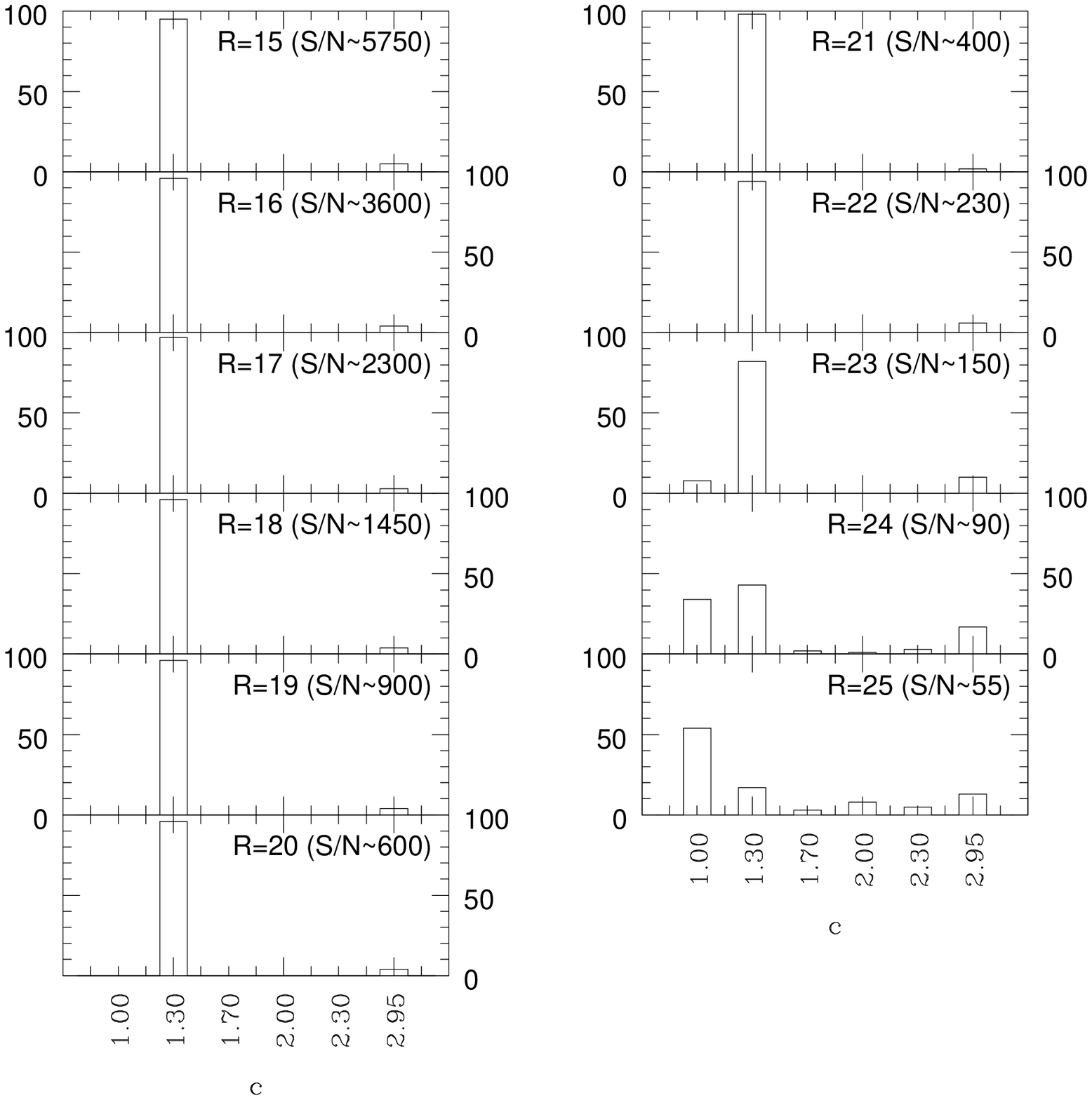}
\label{fig:psferr2}
\end{figure}

To determine the potential extent to which our best fitting
concentrations may be in error due to variations in the PSF, we
constructed sets of artificial clusters using PSF3 and fit them using
PSF1. To span the range of parameters that we might 
encounter when fitting for radial parameters of globulars, we chose two
models, one of which represents the parameters expected for young
clusters based on our fits ($c = 2$, $r_0 = 0.07$) and the other
predicted parameters for an old cluster at the distance of Virgo ($c=
1.3$, $r_0 = 0.28$). We show the histograms of the best fitting output
concentrations for the young clusters at the distance of NGC 3597 and
the old clusters at the distance of Virgo in Figures
\ref{fig:psferr1} and \ref{fig:psferr2}, respectively. From these
we can see that for clusters such as those in our observations, an
inaccurate PSF can lead to systematic errors in the derived
concentration.  Using clusters created with PSF3 and fit with PSF1 we
find that we would end up selecting a larger concentration for our
artificial young clusters at the distance of NGC 3597 (input $c = 2$,
output $c = 2.3$). However, for old clusters at Virgo, which are more
resolved, we find little difference in the output concentration between
clusters created with PSF1 and PSF3 (though the output King radius is
about 8\% higher than the input).  We also note that in Figure
\ref{fig:psferr2}, as in Figures \ref{fig:concerr3} and
\ref{fig:concerr4}, a small number of clusters are seemingly best fit
by large concentrations; as with the other figures, this is 
due to errors in the fits for the smaller concentrations.

\subsection{Variation with Filter Choice}

The F450W filter provides a slightly better discriminant between
different concentrations for identical signal-to-noise values because
of the better resolution provided at shorter wavelengths.
However, in our data sets, for any given magnitude, the F702W
filter has significantly higher counts ($\sim$ 50\%) than the F450W filter,
because of the chosen exposure times and the better quantum efficiency
of the PC chip in the red. For the F450W filter to have a higher
signal-to-noise for equal exposure times, the cluster would need to
have a observed B-R of  $<$ -0.5, which is bluer than even the youngest
clusters which have typical B-R colors of -0.4 based on population
synthesis models (Charlot \& Bruzual 1991). We find for a fixed
exposure time, F702W is preferable to F450W for recovering structural
parameters. This could probably be
generalized to the conclusion that higher signal-to-noise is more
important than the slight increase in spatial resolution obtained from
observing with HST at shorter wavelengths.

\subsection{Discussion}

According to our tests, King radii can be measured to an accuracy of
10-15\% down to a signal-to-noise of $\sim$ 50, assuming that sky noise
is unimportant and that the concentration is well known. Derivation of
the concentration requires substantially higher signal-to-noise because
it becomes critical to measure the wings of the objects. Uncertainties
in the measured concentration as a function of S/N are shown
graphically in Figure \ref{fig:concerr2}, Figure \ref{fig:concerr3},
and Figure \ref{fig:concerr4}. The uncertainty depends on both S/N and
the true concentration of the objects; concentration is harder to
measure accurately for more compact objects.  We find that for young
clusters at a distance of NGC 3597 that the input concentration is
chosen as the best fit output concentration at least 50\% of the time
for S/N $\gtaprx$ 150.  For old clusters at the distance of NGC
3597, the input concentration for the artificial clusters is measured
as the best fit output concentration $\sim$ 50\% of the time only for
S/N $\gtaprx$ 900. For old clusters at Virgo, we find the input concentration as
the best fit concentration at the 50\% level for S/N $\gtaprx$ 100. In each
case there is little scatter in the best fit concentration at these
S/N ratios ($<$ 0.3 in $c$ or less than a factor of two in the ratio of King
to tidal radius), though the scatter quickly increases at fainter
magnitudes.  While these are necessarily only errors for discrete
combinations of parameters, they do provide some insights into the two
most interesting possibilities for the study of globular cluster
structural properties, namely, clusters similar to old Galactic
globulars and clusters similar to massive young protoglobulars.

While our tests were performed for clusters at the distance of NGC
3597, they can provide limits for clusters at other distances.
Globulars in the Virgo cluster, at about 40\% of the distance to NGC
3597, would have a signal-to-noise ratio 2.5 times larger than
identical clusters in NGC 3597 for similar exposure times, assuming
that the dominant source of noise is Poisson noise in the signal. For
an exposure time of 1700 seconds in F702W we could expect to accurately
measure the concentration of an old globular at the distance of Virgo
50\% of the time down to a magnitude of $R= 25$.  At greater distances
than NGC 3597, the situation becomes worse, as the clusters get both
apparently fainter and smaller, and thus we can derive only lower
limits to the spread in parameters. For old globulars at the distance
of NGC 3597, even a cluster as bright as the brightest globular in a
galaxy like M87 would only have an R magnitude of $\sim$ 22,
corresponding to a signal-to-noise ratio of about 230 (for 5000s
exposure time). Thus to accurately determine concentrations for even
the brightest old globulars, one would either take substantially longer
exposures or study closer cluster systems.  As noted above, the fits
are better for apparently larger clusters.  This suggests that
dithering, a way of effectively increasing the resolution, would
improve the quality of the results.

Unfortunately, errors in our understanding of the PSF can lead to
systematic errors in derived radial parameters. Although our simulations
suggest that such errors are not extremely large, they are probably 
responsible for the largest source of uncertainty in our results on
observed clusters, discussed next.

\section{Data}

\subsection{Observations}

We have applied our technique to Hubble Space Telescope Planetary
Camera images of NGC 1275 and NGC 3597. The images of NGC 1275 were
taken on 1995 November 16 and consisted of five exposures of 200, 1000,
and 3 $\times$ 1300 seconds in the F450W filter and five exposures of
200, 900, 1000, and 2 $\times$ 1300 seconds in the F702W filter. The
NGC 3597 images were taken on 1997 May 16 and consisted of four
exposures of 4 $\times$ 1300 seconds in the F450W filter and four
exposures of 1100 and 3 $\times$ 1300 seconds in the F702W filter.
Detailed studies of the blue cluster systems in NGC 1275 and NGC 3597
have been presented in Carlson \etal (1998) and Carlson \etal (1999),
respectively.

For our sample of clusters we have chosen 146 objects from the PC field
of NGC 3597 and 255 objects from the PC field of NGC 1275.  These
subsamples were selected from the list of all clusters in the PC with B
$<$ 26.  All objects with bright neighbors or rapidly varying galaxy
background levels were rejected. This rejection was based on visual
inspection of the images and the light profiles of the candidate
objects.  Objects in the WFs were not included because the poorer
resolution of these frames would make fitting for radial parameters
more difficult.

\subsection{Fitting Radius}

\begin{figure}
\figurenum{8}
\caption{Best fit King radius for a $c= 2$ King model vs Fitting Radius
for 21 of the brightest clusters. The object with R $=$ 19.13 is
likely a foreground star.}
\plotone{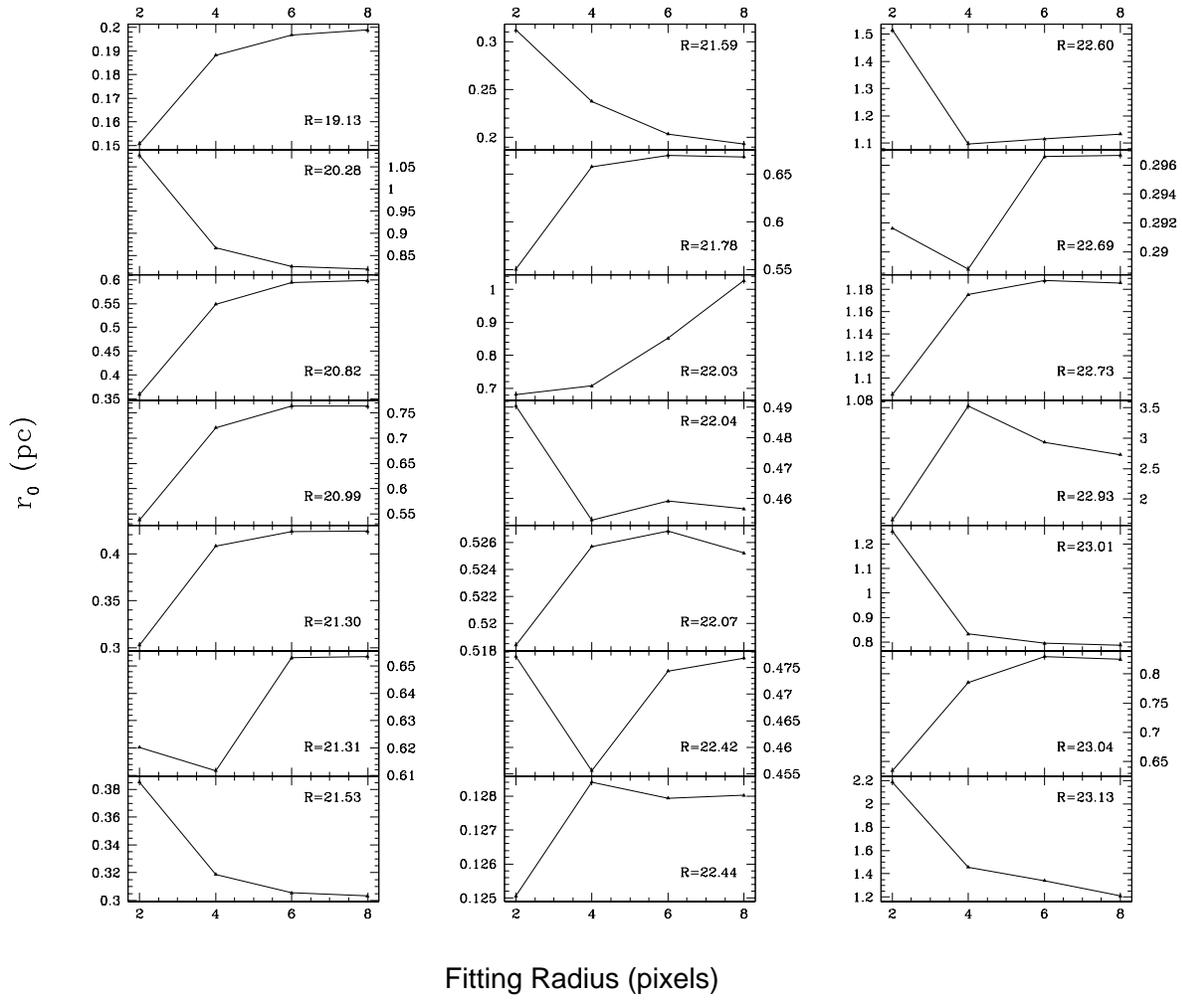}
\label{fig:fitrad}
\end{figure}

The output radial parameters from our fitting program are somewhat
sensitive to our choice of fitting radius. We believe
that this is a result of inaccuracies in the assumed PSF at small radii.
We attempted fits of the clusters using 4 different apertures of 2, 4, 6, and
8 pixels in radius. Figure \ref{fig:fitrad} shows the output King radii for 21 
of the brightest clusters as a function of the different fitting radii.
The variation of the output radial parameter is less than 50\% for all
but three of the clusters.  The 2 pixel aperture varies to the largest
extent, with the greatest deviation between the two and eight pixel
aperture being 81\%. This supports the hypothesis that the problem arises
from PSF inaccuracies at small radii.
The biggest deviation between
the fit for a four pixel aperture and an eight pixel aperture is only
31\%. We adopted 8 pixels as our final fitting radius; note that an 8 pixel
radius includes $\sim$ 94\% of the light of an ideal point source
(Holtzman \etal 1995). 

\subsection{Accuracy of the Fits}

\begin{figure}
\figurenum{9}
\caption{Correspondence of King radius in B and R images for NGC 3597.
All clusters brighter than R $=$ 23 shown as filled circles. Fit was done with
c $=$ 2 King model.}
\plotone{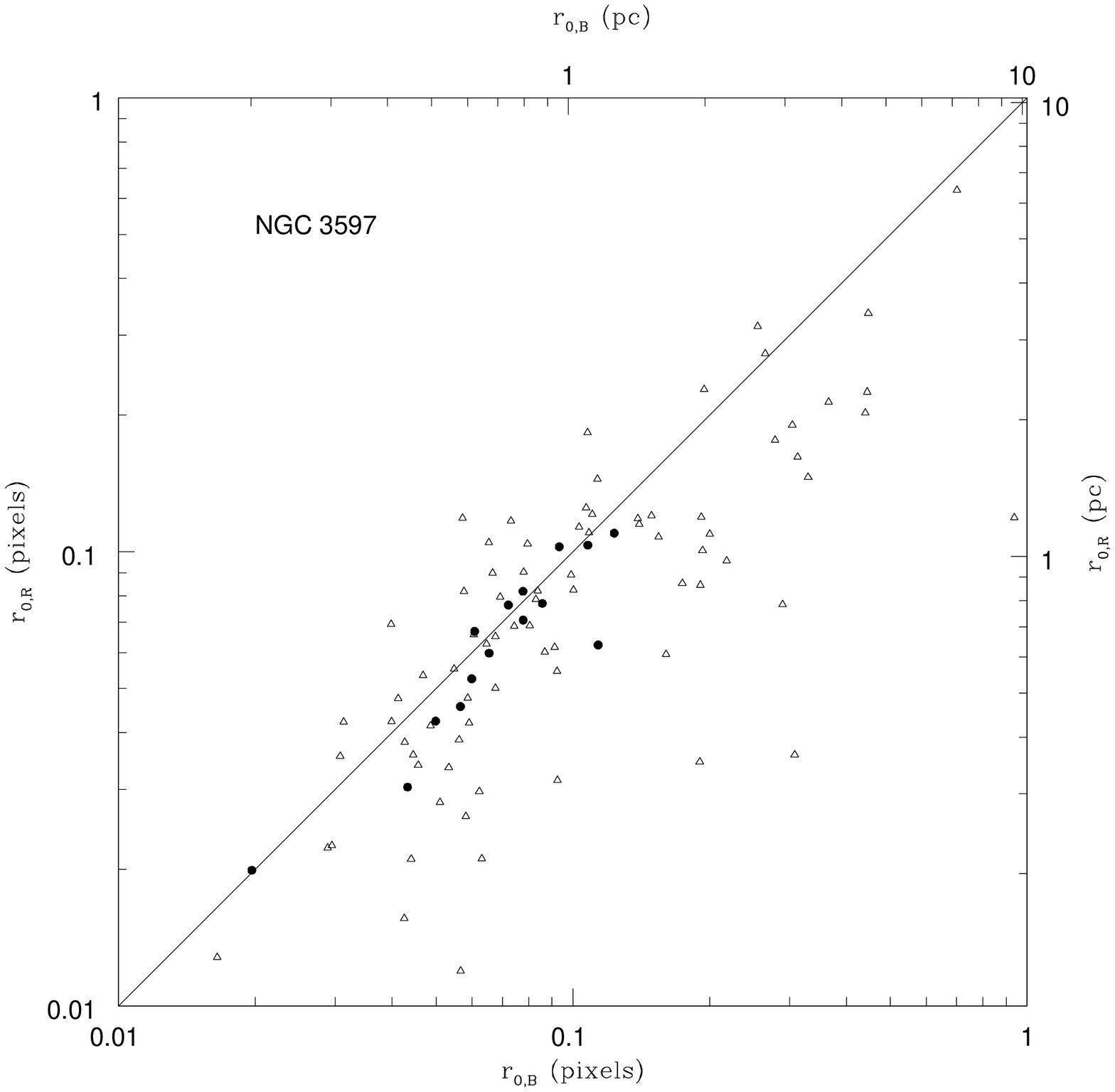}
\label{fig:n3597rvsb}
\end{figure}

\begin{figure}
\figurenum{10}
\caption{Correspondence of King radius in B and R images for NGC 1275.
All clusters brighter than R $=$ 23 shown as filled circles. Fit was done with
c $=$ 2 King model.}
\plotone{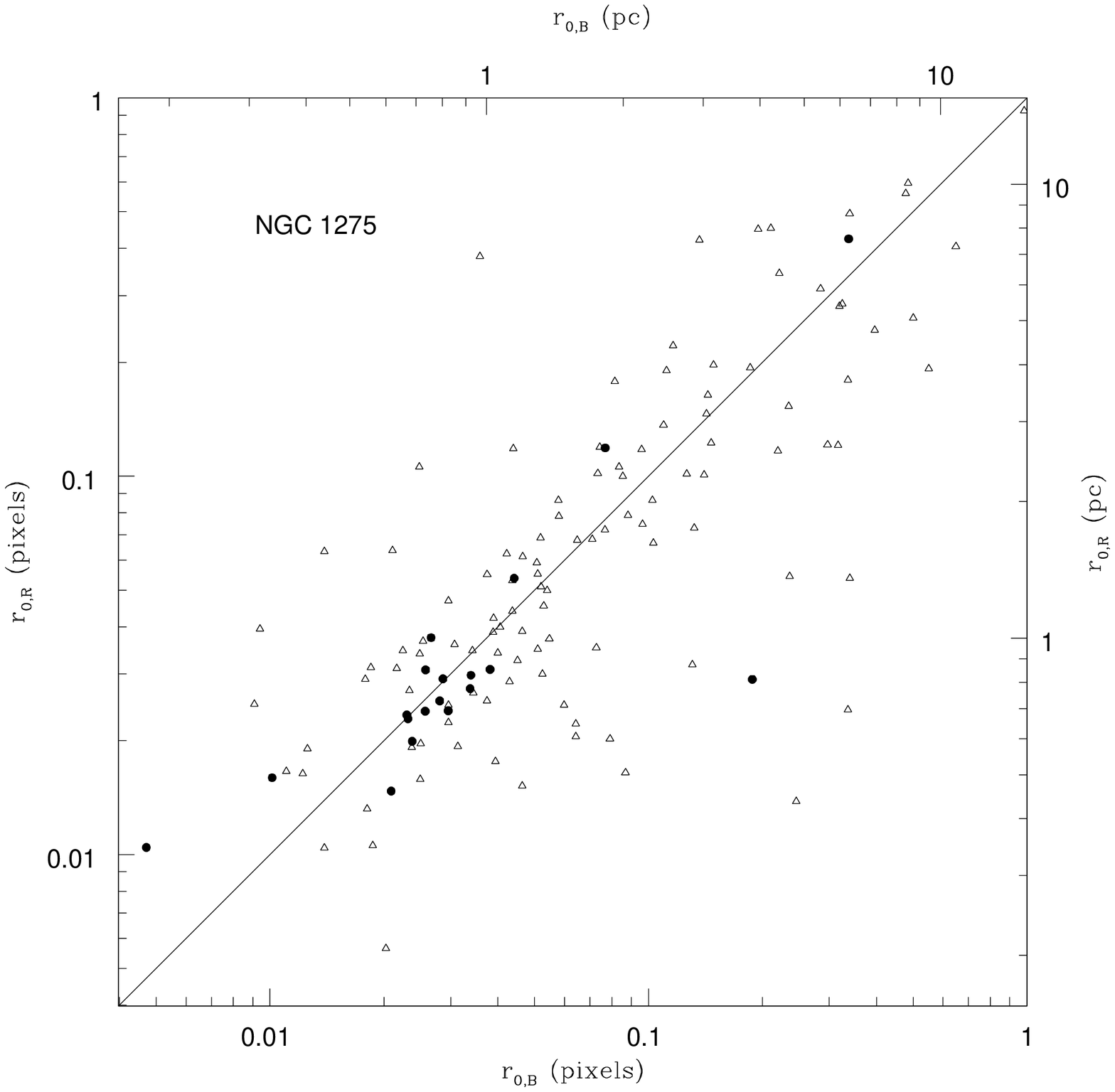}
\label{fig:n1275rvsb}
\end{figure}

Another test of the accuracy of our models is the uniformity of radial
parameters between the two different filters. In Figure \ref{fig:n3597rvsb}
we show the output King radius in R plotted against the output King
radius in B for a King Model ($c = 2$) fit to the clusters
in NGC 3597.  Figure \ref{fig:n1275rvsb} shows an identical plot for NGC
1275. The clusters with B $<$ 23 are shown as filled circles, while the
remaining clusters with B $<$ 26 are shown as open triangles. The solid
line indicates where the clusters would fall if the King radii were
identical in B and R. Note that there is good agreement for the
King radii of the F450W and the F702W images, particularly for the
higher signal-to-noise clusters.  The agreement degrades at fainter
magnitudes, but this is probably due to Poisson noise in the signal and
sky noise for the fainter objects.  One of the brighter clusters in the
NGC 1275 images lies well away from the observed correlation. This
cluster, with a King radius of 0.19 pixels in the B image and 0.03
pixels in the R image is close (within $\sim 12$ pixels) to the two brightest
clusters in our sample and the best fit King radii may
be affected by the wings of those clusters.

\begin{figure}
\figurenum{11}
\caption{Reduced \chisq for three King models (c $=$ 2) vs
R magnitude. Output \chisq connected by lines are from fits to simulations
constructed with the specified PSF and fit with PSF1. PSF1 is a 
sharp focus PSF at the center of the PC image. PSF2 is a poorly focused
PSF. PSF3 is a sharp focus PSF at the edge of the PC field. Values for
observed clusters in NGC 3597 are shown as filled circles.}
\plotone{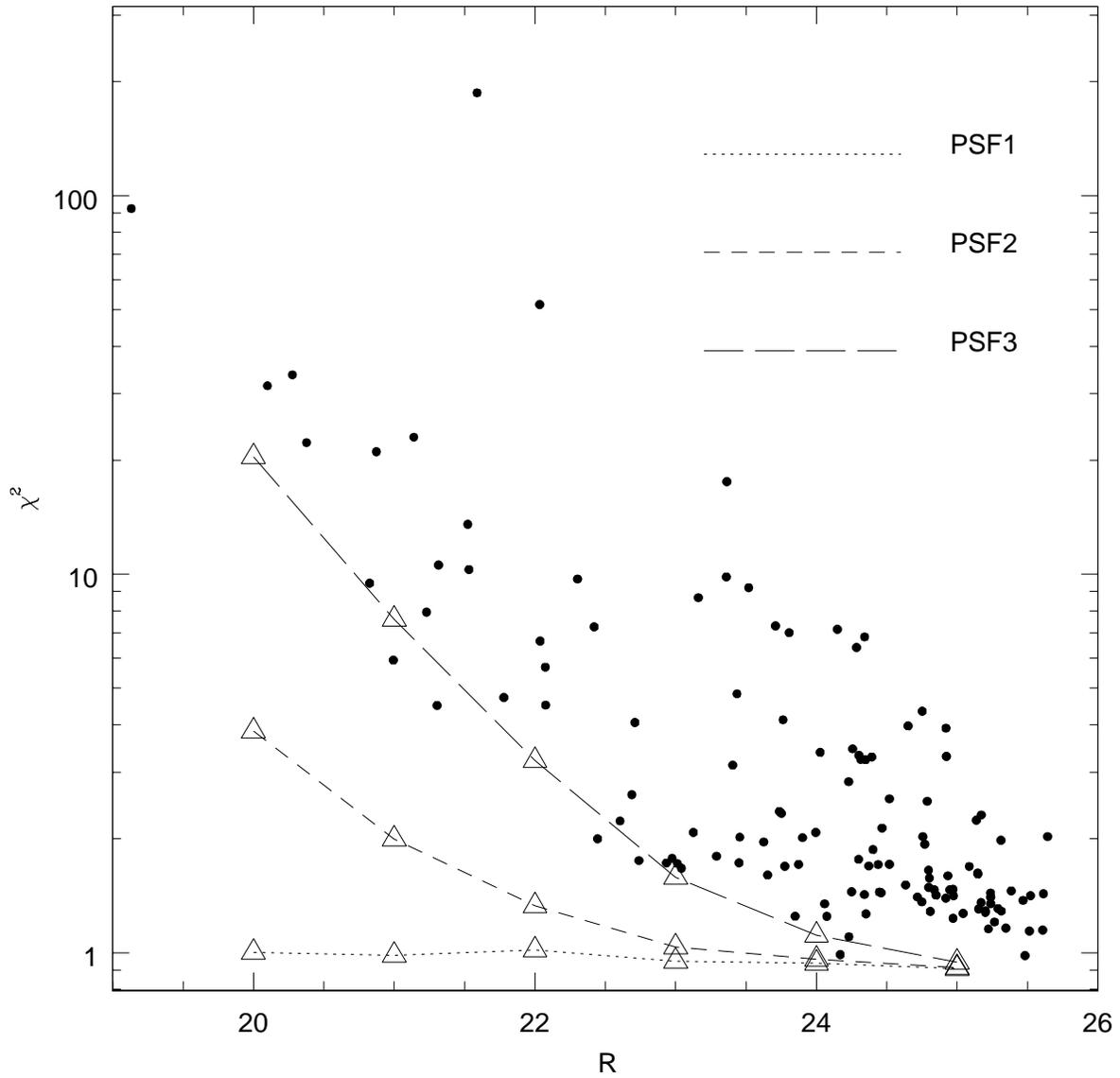}
\label{fig:chi2}
\end{figure}

If our models for the cluster light distribution and PSF are accurate,
we would expect output reduced \chisq values that are close to one.
However, reduced \chisq values for the brightest clusters which are
most sensitive to poor models are dramatically larger than one. In
fact, many of these values are larger than we observed even from the
simulated PSF errors, as shown in Figure \ref{fig:chi2}.  The dominant
source of error may well be problems with the PSF model at small radii.
Evidence to support this comes from Figure \ref{fig:fitrad} where the
fits to the young clusters within two pixels generally deviate more
than any other aperture from the general trends. More direct evidence
from the output \chisq images indicates that the largest part of the
error in the fits comes from within a 2 pixel aperture. Unfortunately,
since the observed \chisq are larger than any of those from the simulations, 
the uncertainties in our results may be larger than we can estimate
from the simulations.

%

\subsection{Results - Radii and Concentrations}

\begin{figure}
\figurenum{12}
\caption{Radial parameters of best fits to the clusters in NGC 3597.
Milky Way King radii shown for comparison.}
\plotone{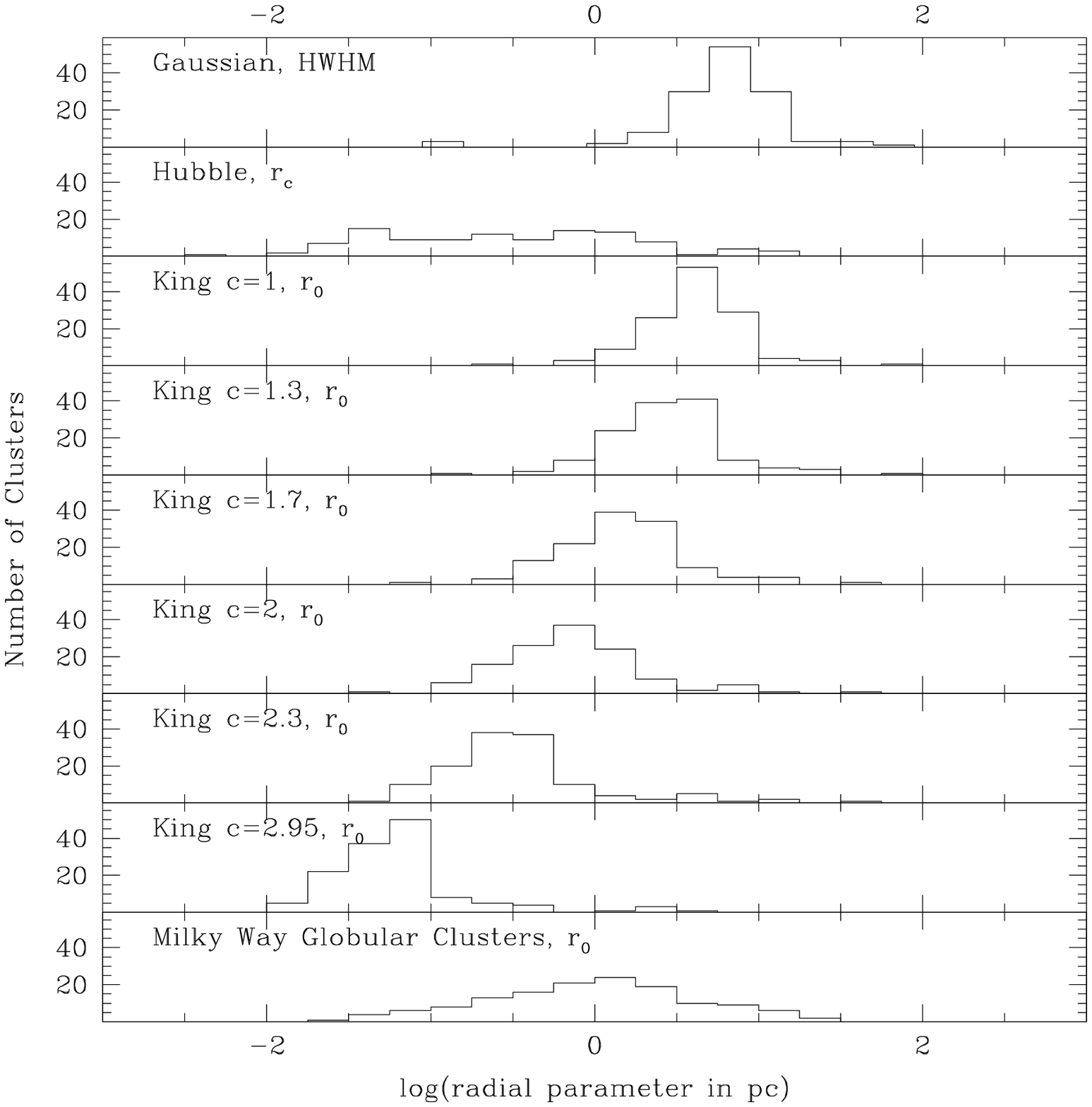}
\label{fig:n3597rad}
\end{figure}

\begin{figure}
\figurenum{13}
\caption{Radial parameters of best fits to the clusters in NGC 1275.
Milky Way King radii shown for comparison.}
\plotone{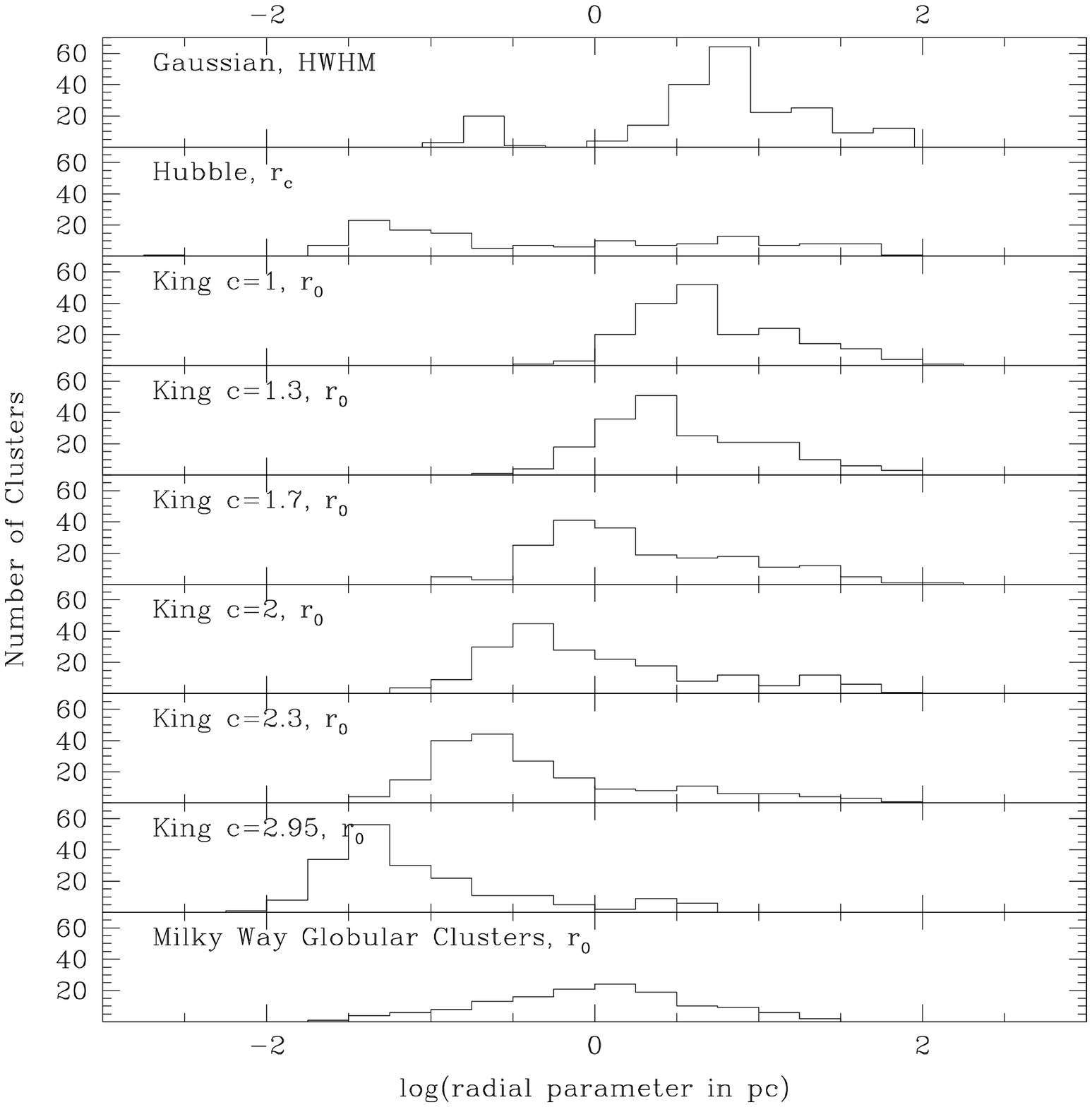}
\label{fig:n1275rad}
\end{figure}

We present histograms of the derived sizes
for 8 different models in Figure \ref{fig:n3597rad} for NGC 3597 and Figure 
\ref{fig:n1275rad} for
NGC 1275. In each plot the top two boxes show the output half width at half 
maximum (HWHM) for the Gaussian
fits and the output core radii for the modified Hubble profiles. The next six
boxes show the output King radii of the clusters assuming they have
concentrations of 1, 1.3, 1.7, 2, 2.3, and 2.95, respectively.
The final box shows the distribution of King radii for Milky Way
globular clusters based on the compilation of Harris (1996).
Note that the values of the King radii for the Galactic globular clusters
are based on simultaneously fitting for both the King radius and the
tidal radius.

\begin{figure}
\figurenum{14}
\caption{Output reduced \chisq vs concentration for 24 of the brightest
clusters. The object with R $=$ 19.13 is likely a foreground star.}
\plotone{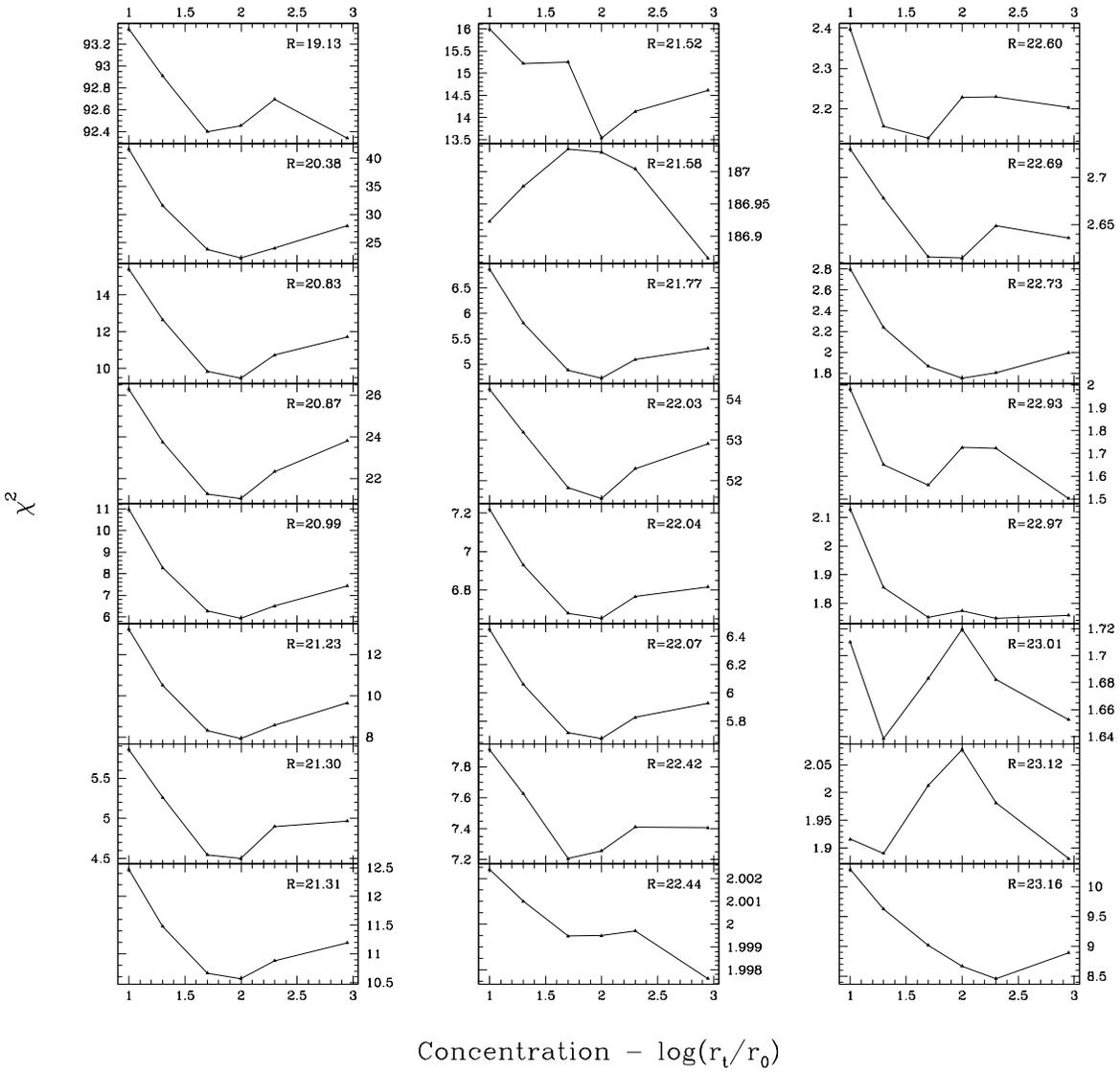}
\label{fig:n3597chi2}
\end{figure}

Results of trying to derive the concentration are shown in
Figure \ref{fig:n3597chi2}, which shows the output reduced \chisq for the 
24 brightest clusters in NGC 3597 as a function of the concentration used
in the model. The brighter clusters appear to be best fit with
concentration of 2, though it is only slightly preferred over 1.7.  Note that
the reduced \chisq values are substantially different from the reduced
\chisq values of the simulations, presumably because the model PSF is not
identical to the actual observed PSF as was the case for the
simulations.  Unfortunately, this allows for the possibility of systematic
errors in our derived parameters.

It is clear from these plots is that we can only determine
concentrations for the brightest clusters in our sample; this is
expected from the simulations given the observed signal-to-noise
ratios of the observed clusters.  For the vast majority of the clusters
in our sample we cannot determine concentration and the King radius
simultaneously.  Based on the derived concentrations of the brightest few
clusters, young blue globular clusters may have concentrations of
$\sim$ 2, corresponding to a ratio of tidal to King radius of 100. The
typical Galactic globular cluster has a concentration of about 1.3 (or
a ratio of 20). Based on these few clusters, it is difficult to
definitively say if young globulars are truly less centrally
concentrated than old globulars.

\begin{figure}
\figurenum{15}
\caption{Histograms of log(Half-light radius in pc) for NGC 3597, NGC 1275,
and the Milky Way.}
\plotone{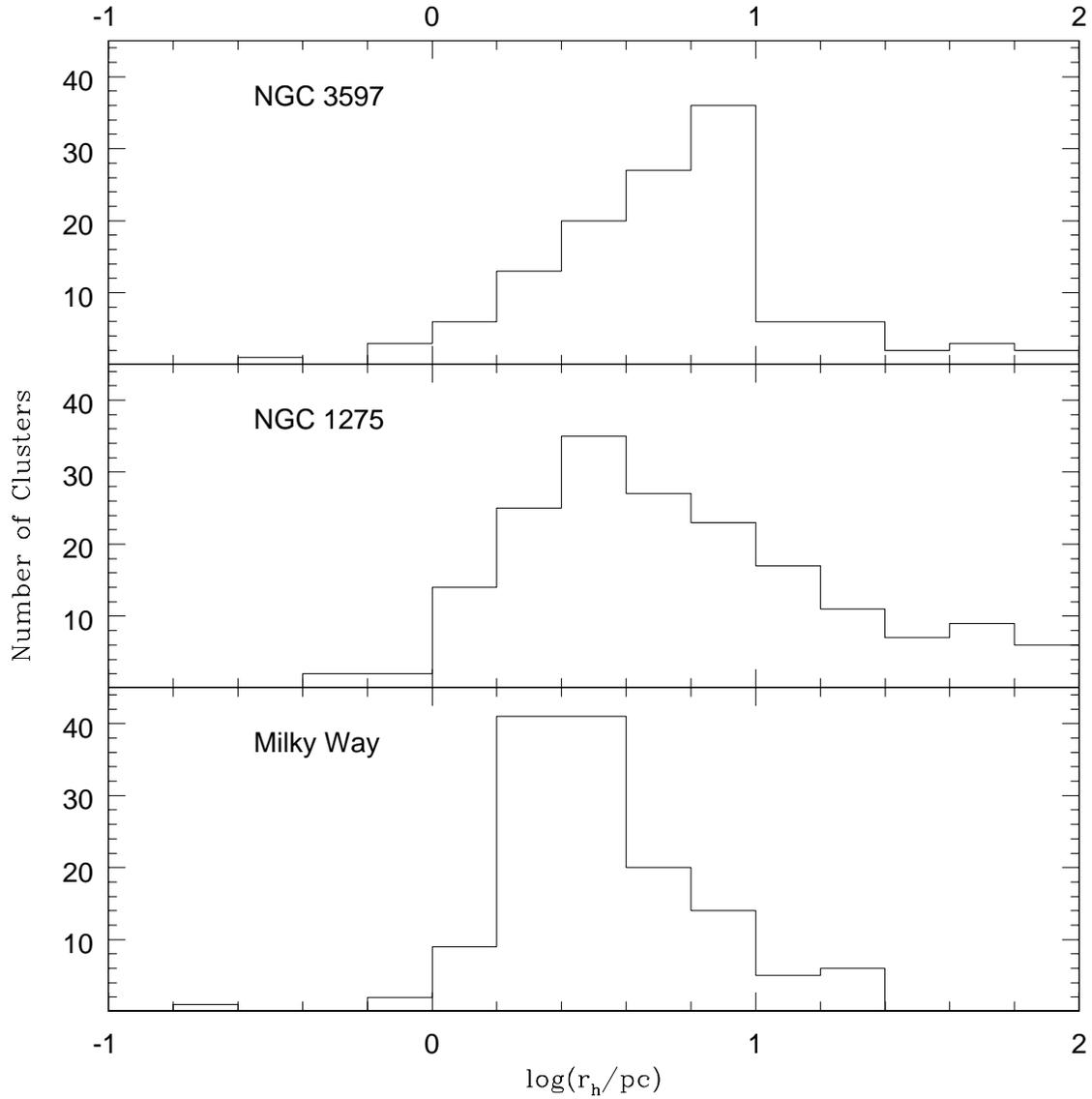}
\label{fig:halflight}
\end{figure}

Since we can only uniquely determine the concentration for a few of the
brightest clusters and the derived value for the King radius is
correlated with the concentration used for the fit, we cannot uniquely
determine King radii for fainter objects. To provide a
model-independent size estimate, we have adopted the solution of Kundu
\& Whitmore (1998) and measured the half-light radius, $r_h$, for the
clusters, which is only weakly dependent on adopted concentration.  We
have determined the half-light radius as a function of King radius using
the $c=2$ King model fit by numerical integration of the
model. Using this result, we convert the best fit King radius 
into a half-light radius.  The use of the half-light radius
simplifies comparison to Galactic globular clusters, and we
present histograms of half-light radii for NGC 3597, NGC 1275, and the
Galactic globular cluster system in Figure \ref{fig:halflight}. We find
the massive young clusters in both NGC 3597 and NGC 1275 have
half-light radii that clearly place them in the same size range as
Galactic globular clusters. The mean sizes may be slightly larger than
those of Galactic globulars; similar results were obtained for the
young clusters in NGC 4038/4039 (Whitmore \etal 1999).

\subsection{Correlations of Sizes with Other Cluster Properties}

\begin{figure}
\figurenum{16}
\caption{Magnitude vs. half-light radius for young clusters in
NGC 3597, NGC 1275, and old globulars in the Milky Way. Apparent magnitudes
in R for the young clusters and absolute magnitudes in V for old
Galactic globulars. Errors on half-light radii are estimated from the
simulations based on S/N and do not include possible systematic errors.
For the faintest clusters, no error bars are shown since they are fainter
than our simulations.} 
\plotone{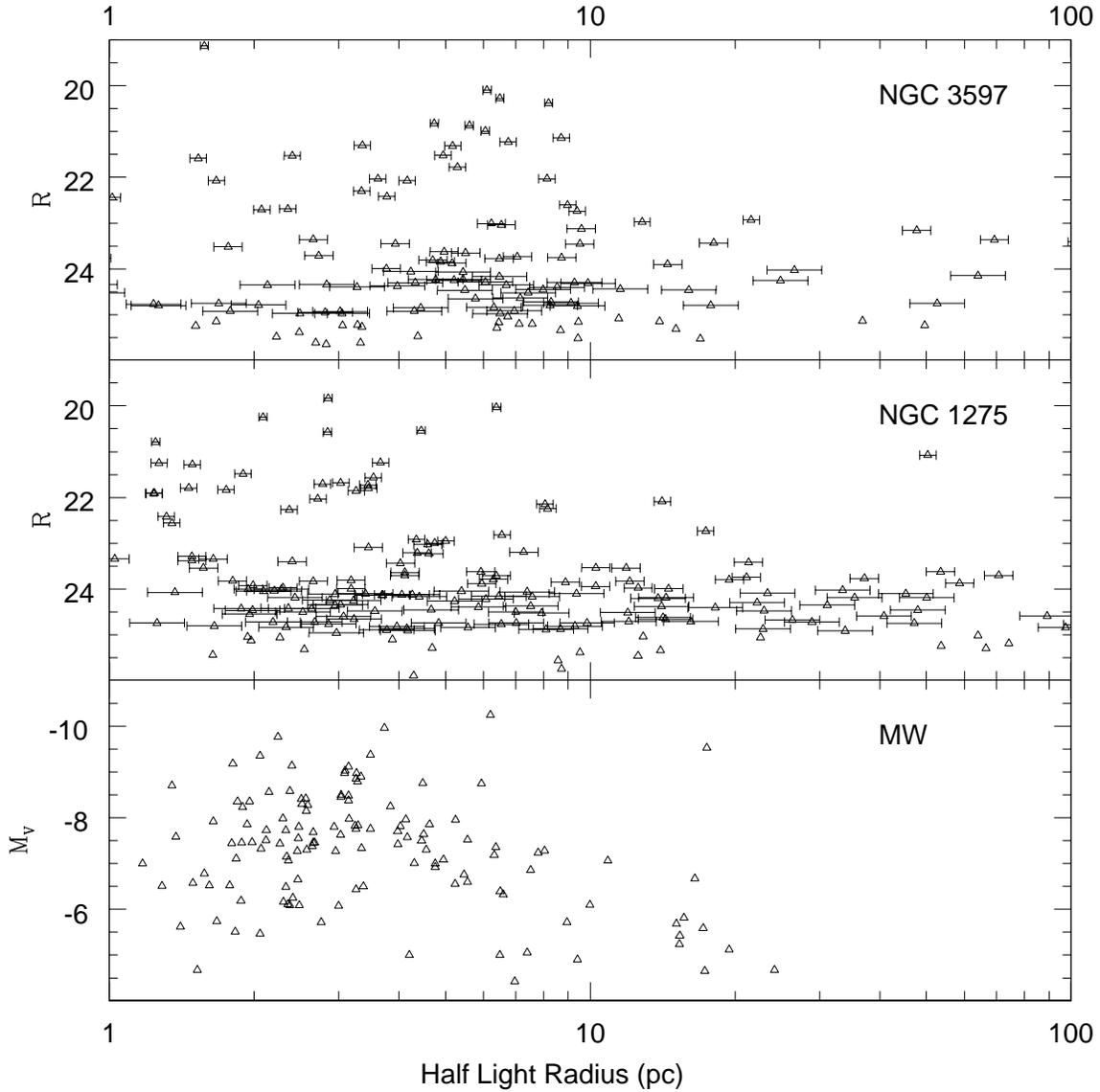}
\label{fig:rvshalflight}
\end{figure}

In Figure \ref{fig:rvshalflight} we display the R magnitude as a
function of half-light radius for the young clusters in NGC 3597 and
NGC 1275. Since most of the clusters in each galaxy have similar 
colors (hence similar stellar populations), the R magnitude is probably
a rough mass indicator. Error bars for the cluster sizes are plotted
using error estimates from our simulations based on the S/N; as noted
above, systematic errors may also exist that we cannot estimate.
While no obvious trend emerges, we do find that the plots
look similar to a plot of M$_V$ versus half-light radius for the
globular clusters in the Milky Way. Clusters with smaller half-light
radii span the whole range of magnitudes, while larger clusters are
preferentially fainter. Similar results were obtained for old globular
cluster systems around NGC 3115 (Kundu \& Whitmore 1998) and M87 (Kundu
\etal 1999).

\begin{figure}
\figurenum{17}
\caption{Distance from the galaxy center in kpc plotted against
half-light radius in pc for NGC 3597, NGC 1275,
and the Milky Way. The twenty brightest clusters in the young cluster
samples are shown as filled circles. Error bars are as in Figure
\ref{fig:rvshalflight}}
\plotone{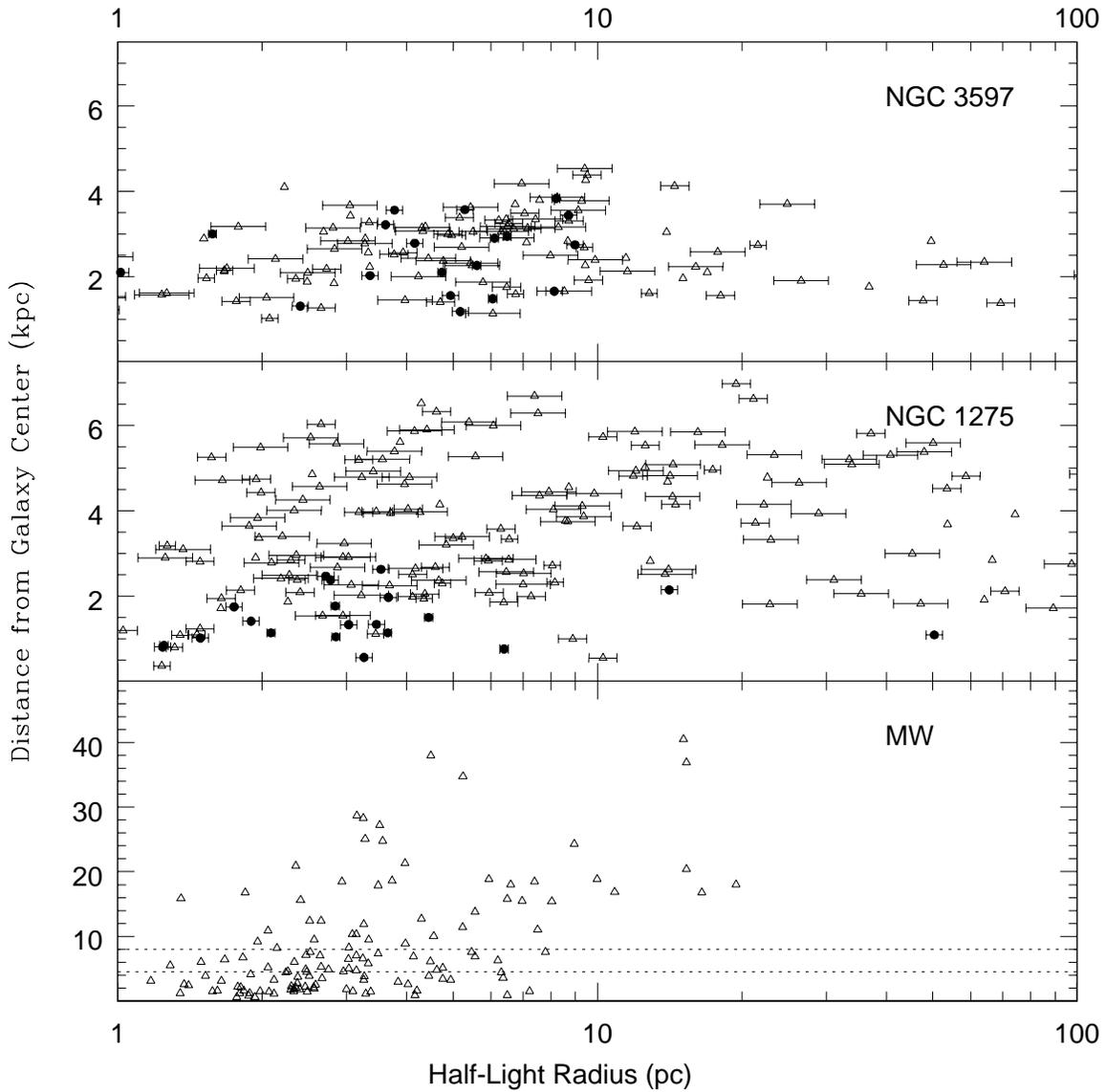}
\label{fig:dvshalflight}
\end{figure}

To investigate the possibility that the size of the young clusters may
be correlated with position in the galaxy, we have plotted the distance
from the galaxy centers against the half-light radii of the clusters in NGC
3597, NGC 1275, and the Milky Way in Figure \ref{fig:dvshalflight}. Note
that clusters in the Galactic sample are plotted to much larger distances
than for NGC 3597 and NGC 1275; the distance to which we can measure
accurate sizes is limited to the size of the PC chip in the WFPC2
observations. The corresponding limiting distances to which we have
observations are shown in the lower panel as dotted lines; we observe
only to $\sim 4.5$ kpc in NGC 3597 and $\sim$ 8 kpc in NGC 1275.
The twenty brightest clusters in each sample
are shown as filled circles. Though there is no correlation between
galactocentric distance and half-light radius, it is notable that there
are no Galactic globulars with half-light radii larger than about 8 pc
within 8 kpc from the Galactic center, though some such clusters exist
at greater distances from the Galactic center. However, these clusters
seem not uncommon in the young cluster samples. Several possibilities
exist to explain this discrepancy. Since we only know the projected
distance from the galaxy centers in our young cluster sample, some
of the clusters may actually be at $>$ 8 kpc. Also, of the
twenty brightest clusters with the best determined half-light radii in
each galaxy, only 2 have half-light radii greater than 10 pc; 
some of the clusters may have larger half-light due to
inaccurate fits due to low signal- to-noise. The possibility also
exists that the clusters within 8 kpc with large half-light radii will
be more prone to disruption by effects such as tidal shocks, and as a result,
we only expect to see them in younger systems.

\subsection{Comparison with Previous Work}

Many previous methods of determining the sizes of young clusters have
relied on measuring aperture magnitude differences.  These magnitude
differences are compared to those for models with similar pixel
centerings to determine a radial parameter.  In this paper, we have
directly fit the surface brightness distributions of the clusters with
the convolution of a PSF and a variety of different
broadening functions. To attempt to better constrain the accuracy of
the various different methods of measuring sizes using aperture
magnitude differences, we compare the results of a few of the aperture
magnitude tests from the literature with our results from direct fits
to the data.

\begin{figure}
\figurenum{18}
\caption{Best fit King radius (c $=$ 2) vs $m_{0.8} - m_3$ for the 100
brightest clusters in NGC 3597. The 20 brightest clusters in this sample
are shown as filled triangles.}
\plotone{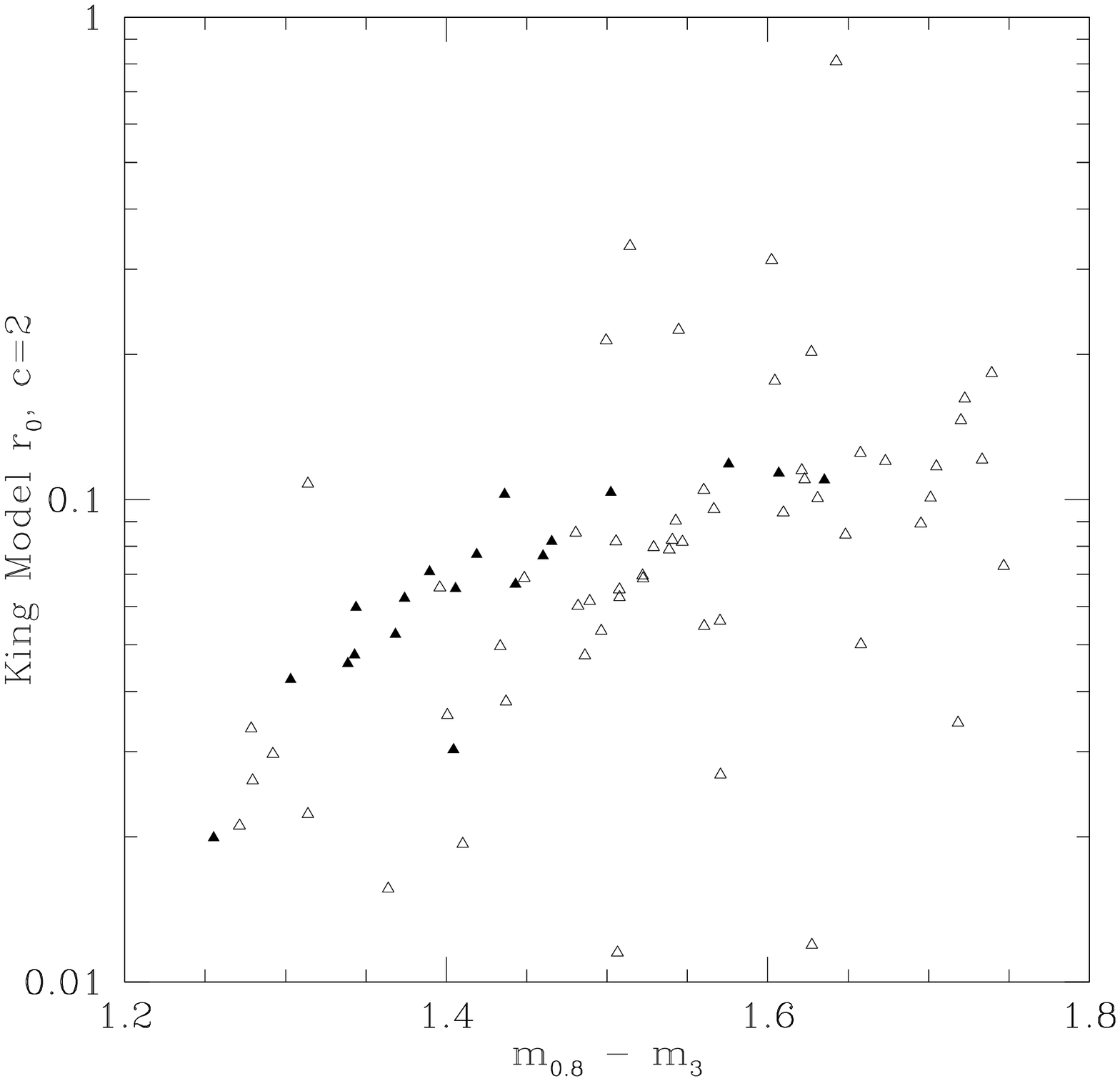}
\label{fig:apsize}
\end{figure}

Several authors have used direct aperture magnitude differences to
determine the structural parameters of young globulars (Whitmore \etal
1993, Whitmore \& Schweizer 1995, Schweizer \etal 1996). The measured
aperture magnitude differences between a 0.5 or 0.8 pixel radius
aperture and a 3 pixel aperture are compared to the same difference for
models of Gaussians convolved with stellar images, which reflect the
PSF.  In Figure \ref{fig:apsize} we show m$_{0.8}$-m$_3$ for the 100 brightest
clusters plotted against the output $r_0$ for the King model fit with
$c=2$ from our fitting program. The general trend confirms
that these models can be used to approximate the structural parameters
of young globulars, but the scatter suggests that there are
potentially large uncertainties judging from the difference in the
results from the different techniques.

\begin{figure}
\figurenum{19}
\caption{Best fit King radius (c $=$ 2) vs $\Delta (m_1 - m_2)$
for the 100 brightest clusters in NGC 3597. The 20 brightest clusters in
this sample are shown as filled triangles.}
\plotone{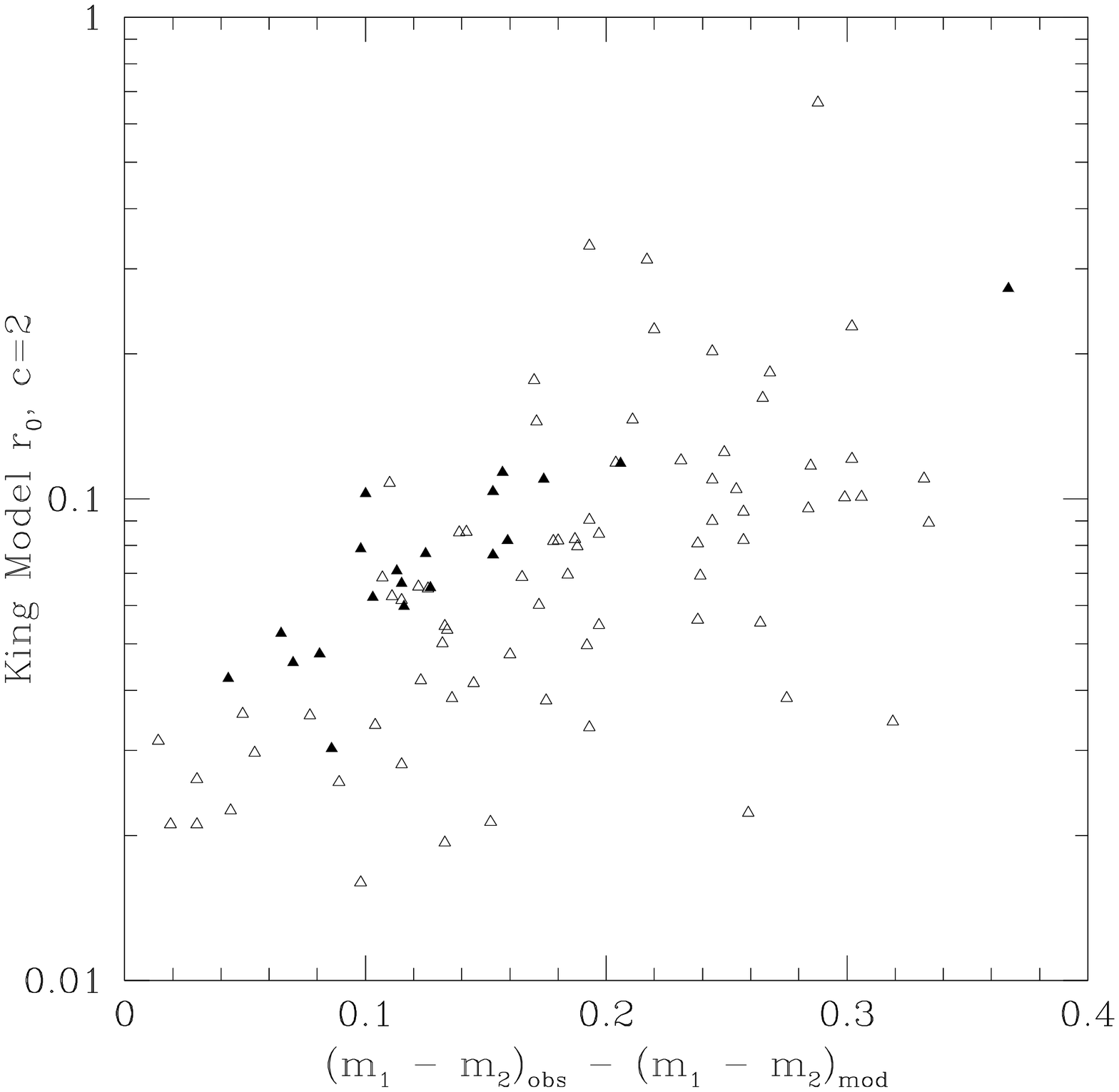}
\label{fig:apsize2}
\end{figure}

A potentially more sophisticated technique originally employed by Holtzman \etal
(1996) uses the aperture magnitude difference between a one and two
pixel aperture magnitude for each cluster. A model PSF is constructed
with identical pixel centering at the same position in the image,
taking into account varying distortions with position. The m$_1$ -
m$_2$ for the model is then subtracted from the observed m$_1$ - m$_2$.
The result is then compared to the same values for modified Hubble
profiles of varying core radii convolved with model PSFs. Using this
technique it was noted that there is some spread in aperture magnitude
differences due primarily to a cluster's centering within a pixel. In
Figure \ref{fig:apsize2} we plot $r_0$ for King model fits with
$c=2$ against the difference in the aperture magnitude
difference between the observed clusters and the PSF. As with the
previous method, there is clearly a general trend, but there is also a
large scatter. In fact, from this comparison, the differential technique 
does not appear to do as well as the simpler aperture difference. It
is possible, however, that some of the scatter actually arises from the
neglect of position dependent PSFs in the \textit{current} fitting method.

\section{Summary}

We have modeled the surface brightness distributions of 146 young
globulars in NGC 3597 and 255 young globulars in NGC 1275 to determine
their structural parameters. We have used Gaussians, modified Hubble
profiles, and King models as the functional forms for the surface
brightness distributions of the underlying clusters and convolved these
with optical models of the Planetary Camera PSF.

From simulations, we find that we can reliably retrieve concentrations
for S/N$\gtaprx$ 600, relatively independently of input
radii/concentrations. If the concentration is known or can be derived,
we find that the measured King radius is accurate to better than 20\%
regardless of the values of the parameters down to S/N $\sim$ 50.  
If we use the wrong concentration for a given artificial cluster, the 
output King radius is systematically wrong. However, if we choose a
concentration, the output King radii from cluster to cluster give a
reasonable approximation of the relative differences in size.  The
half-light radius is relatively independent of chosen concentration,
and is therefore a less model-dependent estimate of cluster sizes.

We find that there is good correspondence between the measured radial
parameters in the F450W and the F702W filters, which suggests that
errors in our PSF models are not color dependent. However, the absolute
errors represented by the output reduced \chisq values of the fits to
the data are larger than might be expected from simple errors in focus
or neglecting the position dependent distortions of the PC field of
view. The suggests that our understanding of the PSF could use some
improvement. PSF inaccuracies probably are responsible for the largest
source of uncertainty in the current results; unfortunately, such errors
could be systematic.

We find preliminary evidence that the young clusters in NGC 3597 are
less centrally concentrated than the old globular clusters in the
Galaxy. This may not be overly surprising, as many forces act to reduce the
tidal radii of globular clusters over their lifetimes, particularly
evaporation.

Comparing the young clusters to old Galactic globulars, we find that
the half-light radius distributions are similar, although the mean
size of the young clusters may be slightly larger; the clusters have
sizes consistent with their being younger versions of Galactic globular
clusters.  The distribution of half-light radii in the young cluster
systems extends to somewhat larger radii than that of the Galaxy. This
may be an evolutionary effect, as larger clusters may be more prone to
disruption.  We also find some relatively large clusters with relatively
small projected galactocentric distance that do not appear to have a
Milky Way counterpart.  There is no strong correlation between magnitude
or galactocentric distance and half-light radius, although there do not
appear to be any very bright clusters that are also large.

Significant future progress will require observations of substantially
higher S/N to allow measurements of concentrations for a larger number
of objects. Spatially dithering of such observations may increase the
accuracy of derived radial parameters. Finally, additional work on improving
the PSF models may be necessary to increase our confidence on derived
cluster parameters.

\acknowledgements

This work was supported in part by NASA under contract
NAS7-918 to JPL and a grant to M. C. from the New Mexico Space
Grant Consortium.

\newpage

\begin{figure}
\figurenum{2}
\plotone{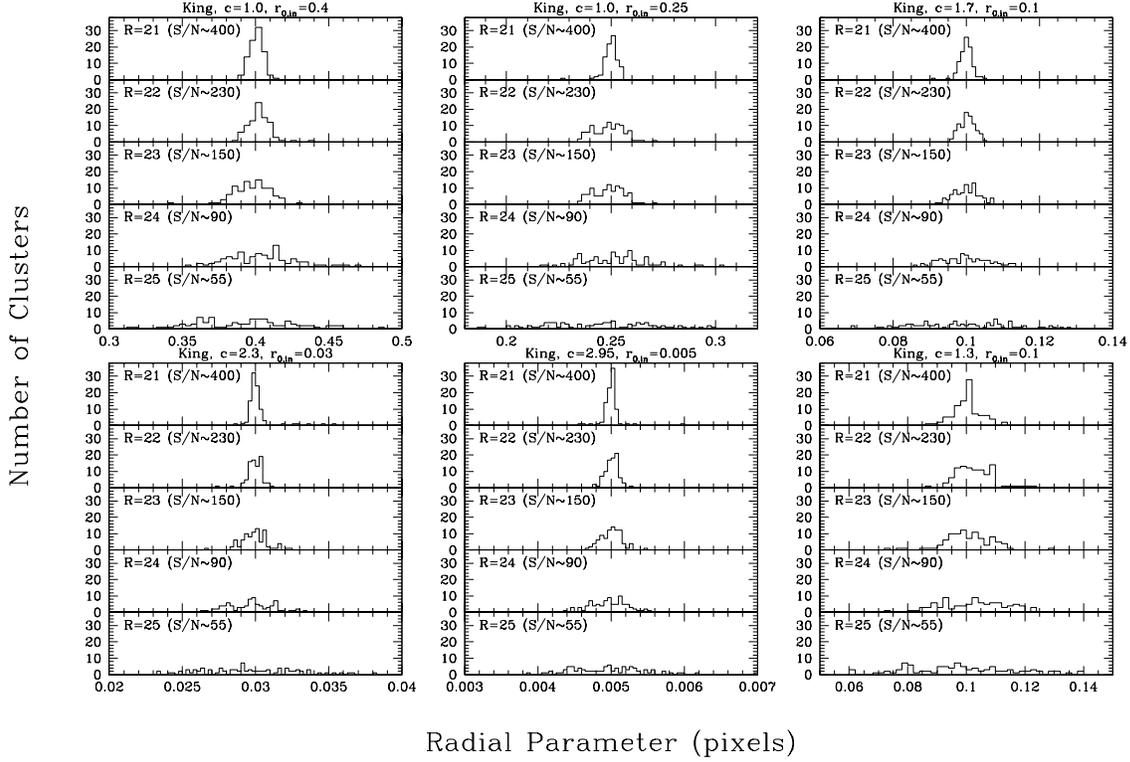}
\label{fig:radialerrb}
\caption{Histograms of output radial parameters from fits to synthetic
clusters. The radial parameter is the FWHM for the Gaussian, the core
radius for the modified Hubble profile, and the King radius for the
King models; all radial units are in pixels.
Input functions and concentrations for the King models were
identical for artificial image construction and fitting. 
}
\end{figure}


\clearpage

\begin{references}

\reference{ashzepf} Ashman, K. M, \& Zepf, S. E., 1998, Globular Cluster
Systems (Cambridge: Cambridge University Press)
\reference{bur95} Burrows, C. J. \etal 1995, Wide Field and Planetary Camera 2 Instrument Handbook, ed. Burows, C. J. (STScI publication), p. 57
\reference{bc91} Charlot, S. \& Bruzual, A. G. 1991, ApJ, 367, 126
\reference{car98} Carlson, M. N. \etal 1998, AJ, 115, 1778
\reference{car99} Carlson, M. N. \etal 1999, AJ, 117, 1700
\reference{chu78} Chun, M. S. 1978, AJ, 83, 1062
\reference{cra85} Crampton, D., Cowley, A. P., Schade, D., \& Chayer, P. 1985, ApJ, 288, 494
\reference{de90} Demers, S., Kunkel, W. E. \& Grondin, L. 1990, PASP, 102, 632
\reference{ef85} Elson, R. A. W. \& Freeman, K. C. 1985, ApJ, 288, 521
\reference{es94} Elson, R. A. W. \& Schade, D. 1994, ApJ, 437, 625
\reference{har96} Harris, W. E. 1996, AJ, 112, 1487
\reference{hol92} Holtzman, J. A. \etal 1992, AJ, 103, 691
\reference{hol95} Holtzman, J. A. \etal 1995, PASP, 107, 156
\reference{hol96} Holtzman, J. A. \etal 1996, AJ, 112, 416
\reference{kin62} King, I. R. 1962, AJ, 67, 471
\reference{kin68} King, I. R., Hedemann, E., Hodge, S. M., \& White, R. E. 1968, AJ, 73, 456
\reference{kon84} Kontizas, M. 1984, A\&A, 131, 58
\reference{kun98} Kundu, A. \&  Whitmore, B. C. 1998, AJ, 116, 2841
\reference{kun99} Kundu, A., Whitmore, B. C., Sparks, W. B., Macchetto, F. D., Zepf, S. E., \& Ashman, K. M. 1999, ApJ, 513, 733
\reference{mar63} Marquardt, D. W. 1963, {\it Journal of the Society for Industrial and Applied Mathematics}, 11, 431
\reference{sch96} Schweizer, F., Miller, B. W., Whitmore, B. C., \& Fall, S. M. 1996, AJ, 112, 1839
\reference{sur93} Surdin, V. G. 1993, in The Globular Cluster Galaxy Connection, ASP Conf. Series Vol. 48, ed. G. H. Smith and J. P. Brodie (San Francisco: BookCrafters, Inc.), p. 342
\reference{wei93} Weinberg, M. D. 1993, in The Globular Cluster Galaxy Connection, ASP Conf. Series Vol. 48, ed. G. H. Smith and J. P. Brodie (San Francisco: BookCrafters, Inc.), p. 689
\reference{whit93} Whitmore, B. C., Schweizer, F., Leitherer, C., Borne, K. \& Robert, C. 1993, AJ, 106, 1354
\reference{ws95} Whitmore, B. C. \& Schweizer, F. 1995, AJ, 109, 960
\reference{whit99} Whitmore, B. C., Zhang, Q., Leitherer, C., Fall, S. M., Schweizer, F., \& Miller, B. W., 1999, AJ, 188, 1551
\reference{zepf} Zepf, S. E., Ashman, K. M, English, J., Freeman, K. C., \& Sharples, R. M, 1999, AJ, 188, 752
\end{references}
\end{document}